\documentclass[journal,twoside]{IEEEtran}
\usepackage{amsmath}
\usepackage{amssymb,amsthm,epsfig,url}
\usepackage{cite}
\usepackage{graphicx}
\newtheorem{theorem}{Theorem}

\newtheorem{property}{Property}

\urldef{\myself}\url{https://sourceforge.net/projects/matlabgammaapproximation/files/Gamma%20Approximation/}

\begin{document}
\title{Analysis of HARQ-IR over Time-Correlated Rayleigh Fading Channels}
\author{Zheng Shi,
        Haichuan~Ding,
        Shaodan Ma,
        and Kam-Weng Tam
\thanks{Manuscript received January 14, 2015; revised May 29, 2015; accepted July 19, 2015. The associate editor coordinating the review of this paper and approving it for publication was M. Elkashlan.}
\thanks{Zheng Shi, Shaodan Ma and Kam-Weng Tam are with the Department of Electrical and Computer Engineering, University of Macau, Macao (e-mail:shizheng0124@gmail.com, shaodanma@umac.mo, kentam@umac.mo).}
\thanks{Haichuan Ding was with University of Macau, and is now with the Department of Electrical and Computer Engineering, University of Florida, U.S.A. (email: dhcbit@gmail.com).}
\thanks{This work was supported by the Research Committee of University of Macau under grants: MYRG078(Y1-L2)-FST12-MSD and MYRG101(Y1-L3)-FST13-MSD.}
}
\markboth{IEEE TRANSACTIONS ON WIRELESS COMMUNICATIONS,~Vol.~xx, No.~x, xxxx~2015}
{Zheng \MakeLowercase{\textit{et al.}}: Analysis of HARQ-IR over Time-Correlated Rayleigh Fading Channels}
\maketitle
\begin{abstract}
In this paper, performance of hybrid automatic repeat request with incremental redundancy (HARQ-IR) over Rayleigh fading channels is investigated. Different from prior analysis, time correlation in the channels is considered. Under time-correlated fading channels, the mutual information in multiple HARQ transmissions is correlated, making the analysis challenging. By using polynomial fitting technique, probability distribution function of the accumulated mutual information is derived. Three meaningful performance metrics including outage probability, average number of transmissions and long term average throughput (LTAT) are then derived in closed-forms. Moreover, diversity order of HARQ-IR is also investigated. It is proved that full diversity can be achieved by HARQ-IR, i.e., the diversity order is equal to the number of transmissions, even under time-correlated fading channels. These analytical results are verified by simulations and enable the evaluation of the impact of various system parameters on the performance. Particularly, the results unveil the negative impact of time correlation on the outage and throughput performance. The results also show that although more transmissions would improve the outage performance, they may not be beneficial to the LTAT when time correlation is high. Optimal rate design to maximize the LTAT is finally discussed and significant LTAT improvement is demonstrated.
\end{abstract}

\IEEEpeerreviewmaketitle
\section{Introduction}
\IEEEPARstart{O}{ver} the last decade, wireless data traffic has experienced an explosive growth. Contrary to this boom in data traffic, transmission reliability and throughput of wireless channels are limited by unideal propagation environment \cite{tse2005fundamentals}. As a combination of forward error control and automatic repeat request (ARQ), hybrid automatic repeat request (HARQ) has been proved as an effective technique to improve transmission reliability and boost system throughput. It thus has been adopted in various wireless standards, such as High Speed Packet Access (HSPA) and Long Term Evolution (LTE) \cite{dahlman20103g, wu2010performance}. Generally, there exist three kinds of HARQ schemes, i.e., Type-I HARQ, HARQ with chase combining (HARQ-CC) and HARQ with incremental redundancy (HARQ-IR). In Type-I HARQ, the erroneously received packets are discarded and each retransmitted packet is decoded independently. In HARQ-CC and HARQ-IR, the previously failed packets are stored and combined with the packets received in subsequent retransmissions for decoding. Specifically, the same packet is retransmitted in each transmission attempt in HARQ-CC scheme, while redundant information is  incrementally transmitted in each HARQ round in the case of HARQ-IR. By exploiting additional coding gain, HARQ-IR is able to achieve a higher link throughput than Type-I HARQ and HARQ-CC \cite{dahlman20103g}. The focus of this paper is thus turned to HARQ-IR scheme.

Performance of HARQ-IR under various wireless systems has been investigated in the literature \cite{wang2010throughput,szczecinski2013rate,makki2014green,makki2012average,larsson2010analysis,Khosravirad2014rate,choi2013energy}. To name a few, delay and throughput of a HARQ-IR enabled multicast system are investigated and scaling laws with respect to the number of users are discovered based on an upper bound of outage probability in \cite{wang2010throughput}. Aiming at throughput maximization, rate allocation and adaptation for HARQ-IR are discussed in \cite{szczecinski2013rate}. For quasi-static fading environments (i.e., channel coefficients are constant during multiple HARQ rounds), power efficiency of HARQ-IR is concerned and optimal power allocation is obtained to minimize outage-limited average transmission power in \cite{makki2014green}. Considering the same quasi-static environments as \cite{makki2014green}, average rate of HARQ-IR enabled spectrum sharing networks is analyzed in \cite{makki2012average}. By expressing outage probability through k-fold convolution, throughput of network coded HARQ-IR with arbitrary number of users is derived in \cite{larsson2010analysis}. In \cite{Khosravirad2014rate}, an optimal rate adaptation policy is proposed for cooperative HARQ-IR with a multi-bit feedback channel. Dynamic programming is then employed to find the optimal rate for maximizing the throughput of the outage-constrained transmission. Moreover, based on dominant term approximation and upper bounds of outage probability, energy-delay-tradeoff (EDT) is analyzed for both one-way and two-way relaying systems with HARQ-IR in \cite{choi2013energy}. Noticing the lack of exact analytical result on outage probability of HARQ-IR, an analytical approach is proposed to derive the outage probability in a closed form through the generalized Fox's H function in \cite{chelli2013performance}. Unfortunately, all of the prior studies are conducted for either quasi-static fading channels or fast fading channels (i.e., channel coefficients in multiple HARQ rounds are independent and identical distributed). The results are not applicable to time-correlated fading channels, which usually occur when the transceiver has low-to-medium mobility \cite{kim2011optimal,jin2011optimal}\footnote{In a dense scattering environment, the time-correlation between two channel amplitudes with time spacing of $\tau$ is quantified by $\rho  = {J_0}^2\left( {2\pi {f_c}\tau v{c^{ - 1}}} \right)$ where $J_0(\cdot)$ denotes the zero-th order Bessel function of the first kind, $f_c$ represents the carrier frequency, $c$ is the speed of light and $v$ refers to the moving speed of a mobile terminal \cite{jakes1994microwave}. Taking a 3GPP LTE system as an example, the successive transmissions are not carried out in adjacent time slots, and the time spacing between two successive HARQ transmissions is $\tau=8$ms \cite{etsi2011lte}. When the LTE system is operated at a carrier frequency of $f_c=2.6$GHz \cite{sesia2011lte}, the time correlation coefficient $\rho$ between the channel amplitudes in two HARQ transmissions is 0.83 and 0.45 for moving speeds of 5~km/h and 10~km/h, respectively.}.

Considering the wide occurrence of time correlation in fading channels in practice, it is necessary and meaningful to analyze the performance of HARQ-IR operating over time-correlated fading channels. However the analysis is very challenging because of the difficulty in handling a product of multiple correlated random variables (RVs). Notice that the analysis is essentially different from \cite{kim2011optimal,jin2011optimal} where HARQ-CC is analyzed and the sum of multiple correlated RVs is concerned. In this paper, we consider HARQ-IR operating over time-correlated Rayleigh fading channels. The accumulated mutual information after multiple HARQ rounds is first expressed as a logarithm function of a product of multiple shifted correlated signal-to-noise ratios. By using polynomial fitting technique, probability distribution function (PDF) of the accumulated mutual information is derived as a product of Gamma distribution and a correction polynomial. Outage probability, average number of transmissions and long term average throughput (LTAT) are then obtained in closed-forms. Moreover, diversity order of HARQ-IR is also analyzed. It is proved that full diversity can be achieved, i.e., the diversity order is equal to the number of transmissions, even under time-correlated fading channels. The impact of channel time correlation on the performance is also investigated and optimal rate design to maximize the throughput is finally discussed. The results reveal that time correlation of the channel causes negative effect on the outage and throughput performance. More HARQ rounds may not be beneficial to the throughput under highly correlated channels, although they do improve the outage performance.

The remainder of this paper is organized as follows. In Section \ref{sec_sys_mod}, a point-to-point HARQ-IR enabled system operating over time-correlated Rayleigh fading channels is introduced. In Section \ref{sec_per}, outage probability, average number of transmission and LTAT are derived in closed-forms by using polynomial fitting technique. Section \ref{sec:dive} analyzes the diversity order of HARQ-IR over time-correlated fading channels. The analytical results are verified through Monte-Carlo simulations, and the impact of time correlation on the performance of HARQ-IR and optimal rate design are then discussed in Section \ref{sec_sim}. Section \ref{sec_con} finally concludes this paper.
\section{System Model}\label{sec_sys_mod}
\begin{figure}
  \centering
  \includegraphics[width=3.5in]{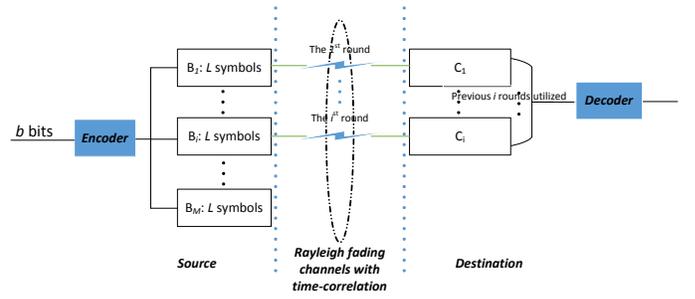}\\
  \caption{System model.}\label{fig_sys_mod}
\end{figure}

Consider a point-to-point system with one source and one destination, as shown in Fig. \ref{fig_sys_mod}. To enhance the transmission reliability, HARQ-IR protocol is adopted here. Notice that most of the prior research on HARQ-IR is carried out over quasi-static or fast fading channels \cite{choi2013energy,chelli2013performance,choi2013onenergy}. Different from the prior analysis, time-correlated fading channels are considered in this paper. Specifically, the HARQ-IR protocol and channel model are introduced in the following.
\subsection{HARQ-IR Protocol}
Following the HARQ-IR protocol, prior to transmit a message with $b$ bits, the source first encodes the message into $M$ packets, each with $L$ symbols. The $M$ packets are denoted as $B_1, B_2, \cdots, B_M$, as shown in Fig. \ref{fig_sys_mod}. Thus the maximum allowable number of transmissions for this message is limited to $M$. Then the source transmits the $M$ packets one by one in multiple HARQ rounds till the destination succeeds to decode the message. If the destination succeeds/fails to decode the message, a positive/negative acknowledgement (ACK/NACK) message will be fed back to the source. At the source, after receiving an NACK message from the destination, the subsequent packet will be delivered in the next HARQ round until the maximum allowable number of transmissions is reached or an ACK message is received. Once either of these two events happens, the source will initiate the transmission of a new message following the same procedure.

At the destination, the received packets are corrupted by fading channels and additive white Gaussian noises. The corrupted packets associated with $B_1, B_2, \cdots, B_M$ are denoted as $\{C_1,C_2,...,C_M\}$. The channel decoder attempts to recover the message based on all the previously received packets. More specifically, after $k$ HARQ rounds, the received packets from $C_1$ to $C_k$ are utilized for decoding the message at the destination. If the message can be successfully recovered, an ACK message will be fed back to the source. Otherwise, a feedback of failure notification will be sent to the source, namely, NACK message.

\subsection{Channel Model}
Denote the modulated signal from the $l$th packet $B_l$ as ${\bf{x}}_l$. The signal received at the destination in the $l$th HARQ round is written as
\begin{equation}
 \label{eqn_received_SD}
 {\bf{y}}_l = h_l{\bf{x}}_l + {\bf{n}}_l
\end{equation}
where ${\bf{n}}_l$ represents complex additive white Gaussian noise with zero mean and variance $\mathfrak{N}_l$, i.e., ${\bf{n}}_l \sim \mathcal{CN}({\bf{0}},\mathfrak{N}_l{{\bf{I}}})$, $h_l$ denotes the block Rayleigh fading channel coefficient in the $l$th HARQ round, i.e., the magnitude of $h_l$ obeys a Rayleigh distribution, such that ${\left| {{h_l}} \right|} \sim Rayleigh({2{\sigma _l}^2})$ and the expectation of the squared channel magnitude is ${\mathrm{E}}\{{{\left| {{h_l}} \right|}^2}\}=2\sigma_l^2$. The PDF of $|h_l|$ is given by
\begin{equation}
{f_{\left| {{h_l}} \right|}}\left( x \right) = \frac{x}{{{\sigma _l}^2}}\exp \left( { - \frac{{{x^2}}}{{2{\sigma _l}^2}}} \right),x \in [0, + \infty ).
\end{equation}

In this paper, time correlation of the channels is considered. Herein, a widely used correlated channel model \cite{beaulieu2011novel} is adopted, that is, the channel coefficients $\left| {\bf{h}} \right|=\{|h_1|, |h_2|, \cdots, |h_K|\}$ are modeled as a multivariate Rayleigh distribution with generalized correlation. The joint PDF of $\left| {\bf{h}} \right|$ follows as \footnote{In fact, the joint PDF of channel amplitudes (\ref{eqn_corr_rayleigh}) is consistent with the conditional PDF of channel power gains given in \cite[Eq. 5]{kim2011optimal}. However, our channel model is different from \cite[Eq. 11]{hu2013optimal} where fast fading channels with imperfect channel state information are considered.}
\begin{align}\label{eqn_corr_rayleigh}
{f_{\left| {\bf{h}} \right|}}\left( {\left| {{h_1}} \right| = {x_1}, \cdots ,\left| {{h_K}} \right| = {x_K}} \right) = \prod\limits_{k = 1}^K {\frac{1}{{{\sigma _k}^2\left( {1 - {\lambda _k}^2} \right)}}}  \notag \\
\times \int\nolimits_{t = 0}^\infty  {{e^{ - \left( {1 + \sum\limits_{k = 1}^K {\frac{{{\lambda _k}^2}}{{1 - {\lambda _k}^2}}} } \right)t}}\prod\limits_{k = 1}^K {{x_k}{e^{ - \frac{{{x_k}^2}}{{2{\sigma _k}^2\left( {1 - {\lambda _k}^2} \right)}}}}}   } \notag \\
\times{}_0{F_1}\left( {;1;\frac{{{x_k}^2{\lambda _k}^2t}}{{2{\sigma _k}^2{{\left( {1 - {\lambda _k}^2} \right)}^2}}}} \right)dt,\, |{\lambda _k}| < 1,
\end{align}
where $_0{F_1}\left( { \cdot } \right)$ denotes the confluent hypergeometric limit function and $\lambda_k$ indicates the correlation degree of the channels. Under this model, the time correlation coefficient between channels associated with the $k$th and the $l$th HARQ rounds is given as
\begin{align}
\label{eqn_cor_fa}
{\rho _{{{{{k}}}},{{{{l}}}}}} &= \frac{{{\rm{E}}\left( {{{\left| {{h_k}} \right|}^2}{{\left| {{h_l}} \right|}^2}} \right) - {\rm{E}}\left( {{{\left| {{h_k}} \right|}^2}} \right){\rm{E}}\left( {{{\left| {{h_l}} \right|}^2}} \right)}}{{\sqrt {{\rm{Var}}\left( {{{\left| {{h_k}} \right|}^2}} \right){\rm{Var}}\left( {{{\left| {{h_l}} \right|}^2}} \right)} }} \notag \\
 &= {\lambda _k}^2{\lambda _l}^2 < 1,\quad k,l \in [1,K],
\end{align}
where the notation ${\rm{Var}}(x)$ denotes the variance of $x$. Notice that this channel model is applicable to the time-correlated channels with correlation coefficients of ${\rho _{k,l}}<1$. For fully
correlated Rayleigh fading channels (i.e., quasi-static Rayleigh fading channels), the channel remains static in all HARQ rounds, i.e., $|h_1| = |h_2| =\cdots=|h_M| \sim Rayleigh(2 {\sigma}^2)$ and  $\rho_{k,l} = 1$. The analysis of HARQ-IR over fully correlated fading channels has been conducted in \cite{makki2014green,makki2012average} and our analysis focuses on the time-correlated channels with ${\rho _{k,l}}<1$.

From (\ref{eqn_received_SD}), the received signal-to-noise ratio (SNR) at the destination in the $l$th HARQ round is written as
\begin{equation}
 \label{eqn_SNR_SD}
\gamma _l = \frac{{{{\left| {h_l} \right|}^2}P_l}}{{\mathfrak{N}_l}}
\end{equation}
where $P_l$ is the transmitted signal power in the $l$th HARQ round. The accumulated mutual information at the destination after $K$ HARQ rounds is then given by
\begin{equation}
I_K^{IR} = \sum\limits_{l = 1}^K {{I_l}}
\end{equation}
where $I_{l}$ represents the mutual information acquired from the $l$th HARQ round and is given as
\begin{equation}
\label{eqn_multual_inf_SD}
{I_{l}} = {\log _2}\left( {1 + \gamma _l} \right).
\end{equation}

\section{Performance Analysis}
\label{sec_per}
\subsection{Performance Metrics}
To investigate the performance of HARQ-IR over time-correlated Rayleigh fading channels, three widely adopted metrics including outage probability, average number of transmissions and long term average throughput (LTAT) are discussed here.

\subsubsection{Outage Probability}
In each HARQ round, the destination combines the current received packet with all the previously received packets for joint decoding. When the accumulated mutual information at the destination is less than the transmission rate $\mathcal{R}$, an outage (i.e., the failure of the decoding) would occur. The outage probability after $K$ HARQ rounds $P_{out}^{IR}\left( K \right)$ is then given as
\begin{equation}\label{eqn:def_out}
P_{out}^{IR}\left( K \right) = \Pr \left( {I_K^{IR} < \mathcal{R}} \right).
\end{equation}
This outage probability can well approximate the error probability when a capacity achieving coding is adopted, and is of great importance in the analysis of HARQ schemes \cite{chelli2013performance}.

\subsubsection{Average Number of Transmissions}
HARQ scheme is a combination of forward error control and automatic repeat request. To enhance the transmission reliability, each message may be retransmitted through multiple HARQ rounds. When the channel condition is good, few retransmissions are sufficient for successful decoding, while more retransmissions are needed over a poor channel. From statistical point of view, it is meaningful to know the average transmission time for each message, which can been well characterized by the average number of transmissions. Given the maximum allowable number of transmissions $M$, the average number of transmissions $\bar {\mathcal{N}} $ is expressed as \cite{chelli2011performance}
\begin{equation}\label{eqn_mean_trans_num}
\bar {\mathcal{N}}  = 1 + \sum\limits_{K = 1}^{M - 1} {P_{out}^{IR}\left( K \right)}.
\end{equation}

\subsubsection{LTAT}
As an effective metric to characterize the system throughput of HARQ schemes, the LTAT given the transmission rate $\mathcal{R}$ and the maximum number of transmissions $M$ is defined as \cite{Chelli2014performance}
\begin{equation}\label{eqn_ltat_asym}
\bar {\mathcal{R}} = \frac{{\mathcal{R}\left( {1 - P_{out}^{IR}\left( M \right)} \right)}}{{\bar {\mathcal{N}} }}.
\end{equation}

Clearly, the average number of transmissions and LTAT only depend on the outage probability, when the transmission rate and the maximum number of transmissions are given. Meanwhile, the outage probability is equivalent to the cumulative distribution function (CDF) ${{F_{I_K^{IR}}}\left( \mathcal R \right)}$ of the accumulated mutual information $I_K^{IR}$, i.e., $P_{out}^{IR}\left( K \right) = {{F_{I_K^{IR}}}\left( \mathcal R \right)}$. As further proved in \cite{makki2014performance}, the essential parameter to characterize the performance of HARQ schemes is the CDF of the accumulated mutual information in each round. It will then be particularly investigated in the next subsection.

\subsection{Analysis of Outage Probability}
By substituting (\ref{eqn_SNR_SD}) and (\ref{eqn_multual_inf_SD}) into (\ref{eqn:def_out}), the outage probability becomes
\begin{align}\label{eqn_out_def}
P_{out}^{IR}\left( K \right) &= \Pr \left( {{{\log }_2}\left( {\prod\limits_{l = 1}^K {\left( {1 + {\gamma_l}} \right)} } \right) < {\cal R}} \right) \notag \\
&= \int_0^{\cal R} {{f_{I_K^{IR}}}\left( x \right)dx}
\end{align}
where ${{f_{I_K^{IR}}}\left( x \right)}$ stands for the PDF of $I_K^{IR}$. Since the considered channel is Rayleigh fading, it is easy to get that the received SNR ${\gamma_l}$ follows exponential distribution with a PDF of
\begin{equation}\label{eqn_Xl_pdf}
{f_{{\gamma_l}}}\left( x \right) = \frac{1}{{{2\sigma _l'}{^2}}}\exp \left( { - \frac{{{x}}}{{2{\sigma _l'}{^2}}}} \right),x \in [0, + \infty )
\end{equation}
where ${\sigma _l'} = {\left( {{{{P_l}}}/{{{\mathfrak{N}_l}}}} \right)^{\frac{1}{2}}}{\sigma _l}$. Due to the time correlation in the channels, the SNRs ${\gamma_l}$ are correlated. By making simple substitutions of variables on (\ref{eqn_corr_rayleigh}), the joint distribution of ${\boldsymbol \gamma_{1:K}} = \left\{ {{\gamma_1},{\gamma_2}, \cdots ,{\gamma_K}} \right\}$ can be readily derived as
\begin{align}\label{eqn_joint_pdf_X}
{{f_{\boldsymbol \gamma_{1:K}}}\left( {{\gamma _1} = {x_1}, \cdots ,{\gamma _K} = {x_K}} \right) = \prod\limits_{k = 1}^K {\frac{1}{{2{\sigma _k'}{^2}\left( {1 - {\lambda _k}^2} \right)}}}   } \notag\\
\times {\int\nolimits_{t = 0}^\infty  {{{\rm{e}}^{ - \left( {1 + \sum\limits_{k = 1}^K {\frac{{{\lambda _k}^2}}{{1 - {\lambda _k}^2}}} } \right)t}}\prod\limits_{k = 1}^K {{e^{ { - \frac{{{x_k}}}{{2{\sigma _k'}{^2}\left( {1 - {\lambda _k}^2} \right)}}} }}} } } \notag\\
\times{}_0{F_1}\left( {;1;\frac{{{x_k}{\lambda _k}^2t}}{{2{\sigma _k'}{^2}{{\left( {1 - {\lambda _k}^2} \right)}^2}}}} \right) dt.
\end{align}
It is clear from (\ref{eqn_out_def}) that the distribution of a product of time-correlated shifted-exponential RVs is necessary to derive the outage probability. As reported in the literature, Mellin transform can be exploited to derive the distribution of the product of independent RVs \cite{springer1966distribution,lomnicki1967distribution,springer1970distribution,salo2006distribution,yilmaz2009product,yilmaz2010outage,chelli2013performance}. Unfortunately, it is inapplicable for the case with correlated RVs due to the involvement of multiple integral. In fact, the presence of time correlation makes the derivation of the exact distribution of ${I_K^{IR}}$ intractable. To proceed with the analysis, we resort to find a good approximation of the distribution of ${I_K^{IR}}$ based on polynomial fitting technique which will be introduced in the following.

As shown in \cite{Chelli2014performance}, the accumulated mutual information in HARQ-IR systems over \textit{independent} fading channels can be well approximated as a Gamma RV by using
Laguerre series. Inspired by this result, the PDF of the accumulated mutual information $I_K^{IR}$ over \textit{correlated} Rayleigh fading channels can be written as the product of a Gamma PDF $\varphi(x)$ and a correction term $\psi \left( x \right)$ as \footnote{Notice that when the channels are independent fading, the correction term can be approximated as $\psi \left( x \right) \approx 1$ and the PDF is reduced as ${f_{I_K^{IR}}}(x) \approx \varphi (x)$ \cite{Chelli2014performance}.}
\begin{equation}\label{eqn:relation_f_phi}
{f_{I_K^{IR}}}(x) = \varphi (x)\psi \left( x \right).
\end{equation}
In (\ref{eqn:relation_f_phi}), the Gamma PDF $\varphi (x)$ serves as a basis function and is given by
\begin{equation}\label{eqn_log_gamma}
\varphi(x) = \frac{{{x^{\zeta - 1}}{e^{ - \frac{x}{\theta }}}}}{{{\theta ^\zeta}\Gamma \left( \zeta \right)}}, x \ge 0
\end{equation}
where $\Gamma (\cdot)$ denotes Gamma function, and the parameters $\zeta$ and $\theta$ are determined by matching the first two moments of $f_{I_K^{IR}}(x)$ with that of $\varphi(x)$, thus leading to \cite{primak2005stochastic}
\begin{equation}\label{eqn_k}
\zeta = \frac{{{{\mathcal M}^2}\left( 1 \right)}}{{{\mathcal M}\left( 2 \right) - {{\mathcal M}^2}\left( 1 \right)}}
\end{equation}
\begin{equation}\label{eqn_theta}
\theta  = \frac{{{\mathcal M}\left( 2 \right) - {{\mathcal M}^2}\left( 1 \right)}}{{{\mathcal M}\left( 1 \right)}}
\end{equation}
where ${\mathcal M}(i)$ denotes the $i$th moment with respect to $f_{I_K^{IR}}(x)$. As shown in Appendix \ref{dm}, the $i$th moment can be derived using Gaussian Quadrature as
\begin{align}\label{eqn_M_fin_sec}
{\mathcal M}\left( i \right) \approx \frac{{i!}}{{1 + \sum\limits_{k = 1}^K {\frac{{{\lambda _k}^2}}{{1 - {\lambda _k}^2}}} }}\sum\limits_{\sum\limits_{l=1}^K{i_l}=i,{i_l} \ge 0} {\frac{1}{{{i_1}!{i_2}! \cdots {i_K}!}} } \notag\\
 \times \sum\limits_{{q_k} \in \left[ {1,{N_Q}} \right],k \in \left[ {1,K} \right]} {\prod\limits_{k = 1}^K {{\varrho _{{q_k}}}{{\log }_2}^{{i_k}}\left( {1 + {w_k}{\xi _{{q_k}}}} \right)} } \notag \\
 \times \Psi _2^{\left( K \right)}\left( {1;1,1, \cdots ,1;{\varpi _1}{\xi _{{q_1}}},{\varpi _2}{\xi _{{q_2}}}, \cdots ,{\varpi _K}{\xi _{{q_K}}}} \right).
\end{align}
where ${\varpi _k} = \frac{{{\lambda _k}^2}}{{1 - {\lambda _k}^2}}{\left( {1 + \sum\nolimits_{l = 1}^K {\frac{{{\lambda _l}^2}}{{1 - {\lambda _l}^2}}} } \right)^{ - 1}}$, ${w_k} = 2{\sigma _k'}^2\left( {1 - {\lambda _k}^2} \right)
$, $N_Q$ is the quadrature order, the weights $\varrho _p$ and the abscissas ${{\xi _p}}$ are tabulated in \cite[Table 25.9]{abramowitz1972handbook}, and $\Psi _2^{\left( K \right)}(\cdot)$ is defined as the confluent form of Lauricella hypergeometric function \cite[Definition A.20]{mathai2009h}, \cite{saigo1992some}. Specifically, ${\mathcal M}(0)=1$.

On the other hand, the correction term $\psi \left( x \right)$ in (\ref{eqn:relation_f_phi}) is used to compensate the difference between ${f_{I_K^{IR}}}(x)$ and the basis function $\varphi (x)$. Apparently, it is very difficult to derive the exact expression for $\psi \left( x \right)$. However, $\psi \left( x \right)$ can be generally approximated as a polynomial $\hat \psi_N(x) \in \mathbb{P}_N$ with degree $N$ by means of polynomial fitting technique \cite{Germund2008numerical}, where the fitting error is characterized as $e \left( x \right)= \psi \left( x \right) - \hat \psi_N(x)$. The remaining problem here is then to find the optimal polynomial $\hat \psi_N(x)$ which can minimize the mean square error (MSE), i.e., ${\rm{E}}\{ {e\left( x \right)^2}\} {\rm{ = }}\int_{ 0 }^\infty  {\varphi \left( x \right){{\left( { \psi \left( x \right) - \hat \psi_N(x)} \right)}^2}dx} $.

It is widely known that any polynomial can be written as a unique linear combination of orthogonal polynomials. Denote the monic orthogonal polynomials $\boldsymbol{\mathcal{P}}(x) = [\mathcal{P}_0(x),\mathcal{P}_1(x),\cdots,\mathcal{P}_N(x)]^{\rm{T}}$ with respect to a measure $d\mu (x)$ be the basis in the space of polynomials of degree less than or equal to $N$, where ${{\mathcal P}_n}\left( x \right) = \sum\nolimits_{k = 0}^n {{C_{n,k}}{x^k}}  \in {{\mathbb P}_n}$ and $C_{n,n}=1$. As pointed out in \cite[Theorem 1.27]{gautschi2004orthogonal}, $\boldsymbol{\mathcal{P}}(x)$ is uniquely determined given the measure $d\mu (x)$. Moreover, the monic orthogonal polynomials obey the orthogonality
\begin{equation}\label{eqn_dirac_def}
\left\langle {{\mathcal P_n}\left( x \right),{\mathcal P_k}\left( x \right)} \right\rangle  = {\delta _{n,k}}{{\cal D}_n} = \left\{ {\begin{array}{*{20}{c}}
{{{\cal D}_n},}&{n = k};\\
{0,}&{else}.
\end{array}}, \right.
\end{equation}
where $\delta _{n,k}$ indicates Kronecker delta function, $\left\langle {g\left( x \right),h\left( x \right)} \right\rangle $ denotes an inner product defined on $2$-norm Lebesgue space $L^2(\mathbb{R},\mathcal{F},\mu )$ \footnotemark ~with the measure $d\mu (x)$, that is,
\footnotetext{Herein, $(\mathbb{R},\mathcal{F},\mu )$ is a measure space, where $\mathcal{F}$ is $\delta$-algebra over $\mathbb{R}$.}
\begin{equation}\label{eqn_inner_def}
\left\langle {g\left( x \right),h\left( x \right)} \right\rangle  = \int_{ - \infty }^{ + \infty } {g\left( x \right)h(x) d \mu(x)},
\end{equation}
and ${\cal D}_n$ is a non-zero parameter which will be specified later. With the monic orthogonal polynomials $\boldsymbol{\mathcal{P}}(x)$, $\hat \psi_N(x)$ can be written as
\begin{equation}\label{eqn_inter_def}
\hat \psi_N(x) {\rm{ = }}\sum\limits_{i = 0}^N {{\eta _i}{\mathcal{P}_i}\left( x \right)} = {\boldsymbol{\eta}}^{\rm T} \boldsymbol{\mathcal{P}}(x)
\end{equation}
where the column vector ${\boldsymbol{\eta}} = [\eta_0,\eta_1,\cdots,\eta_N]^{\rm{T}}$ can be regarded as the corresponding coordinate vector of $\hat \psi_N(x)$ in the space $\mathbb{P} _N$. Substituting (\ref{eqn_inter_def}) into (\ref{eqn:relation_f_phi}), the PDF ${f_{I_K^{IR}}}(x) $ can be approximated as
\begin{equation}\label{eqn:reulst_PDF}
{f_{I_K^{IR}}}(x){\approx} f_{{{\boldsymbol \gamma_{1:K}}},N}^{IR}(x) = \varphi (x)\hat \psi_N\left( x \right) = \varphi (x){\boldsymbol{\eta}}^{\rm T} \boldsymbol{\mathcal{P}}(x).
\end{equation}

To guarantee the approximation accuracy, we need to find the optimal polynomial correction term $\hat \psi_N(x)$ (i.e., the optimal ${\boldsymbol{\eta}}$ and $\boldsymbol{\mathcal{P}}(x)$) which can minimise the MSE ${\rm{E}}\{ {e\left( x \right)^2}\}$. It is noteworthy that the resulting approximated PDF $f_{{{\boldsymbol \gamma_{1:K}}},N}^{IR}(x)$ should be normalized, such that
\begin{equation}\label{eqn:norm_app_PDF_1}
\int_0^\infty  f_{{{\boldsymbol \gamma_{1:K}}},N}^{IR}(x)dx  = 1.
\end{equation}
Considering this constraint and the fact that the monic orthogonal polynomials $\boldsymbol{\mathcal{P}}(x)$ are uniquely determined by the measure $d\mu (x)$, given the measure $d\mu (x)$ and $N$, the approximation problem can be formulated as
\begin{equation}\label{eqn_op}
\begin{aligned}
& \underset{{\boldsymbol{\eta }}}{\text{min}} & &{{\cal S}_{mse}}({\boldsymbol{\eta }}|d\mu (x),N) \\
& & &= \int\nolimits_{ 0 }^\infty  {\varphi \left( x \right){{\left( {\sum\limits_{i = 0}^N {{\eta _i}{{\cal P}_i}\left( x \right)}  - \psi \left( x \right)} \right)}^2}dx}   \\
& \text{s.t.} & &\int_0^\infty  f_{{{\boldsymbol \gamma_{1:K}}},N}^{IR}(x)dx  = 1\\
\end{aligned}
\end{equation}
This minimization problem can be solved by adopting the method of Lagrange multiplier. The Lagrangian function corresponding to the minimization problem can be written in matrix form as
\begin{align}\label{eqn:lag_fun}
\Lambda ({\boldsymbol{\eta }},\varsigma |d\mu (x),N) &= \int_{ 0 }^\infty  {\varphi \left( x \right){{\left( {\sum\limits_{i = 0}^N {{\eta _i}{{\cal P}_i}\left( x \right)}  - \psi \left( x \right)} \right)}^2}dx} \notag \\
&+ \varsigma \left( {\int_0^\infty  {f_{{{\boldsymbol \gamma_{1:K}}},N}^{IR}(x)dx}  - 1} \right) \notag\\
& = {{\boldsymbol{\eta }}^{\rm{T}}}{\bf{A\boldsymbol{\eta} }} - 2{{\boldsymbol{\eta }}^{\rm{T}}}{\bf{b}} + \varsigma {{\boldsymbol{\eta }}^{\rm{T}}}{\bf{d}} + c - \varsigma
\end{align}
where $\varsigma$ represents the Lagrange multiplier, $\bf A$, $\bf b$ and $\bf d$ are given by (\ref{eqn:def_mtr_A})-(\ref{eqn:def_vec_d}) as shown on the top of next page, respectively,
\begin{figure*}[!t]
\begin{align}\label{eqn:def_mtr_A}
{{\bf{A}} = \left[ {\begin{array}{*{20}{c}}
{\int_{ - \infty }^\infty  {\varphi \left( x \right){{\mathcal P}_0}^2\left( x \right)dx} }&{\int_{ 0 }^\infty  {\varphi \left( x \right){{\mathcal P}_0}\left( x \right){{\mathcal P}_1}\left( x \right)dx} }& \cdots &{\int_{ 0 }^\infty  {\varphi \left( x \right){{\mathcal P}_0}\left( x \right){{\mathcal P}_N}\left( x \right)dx} }\\
{\int_{ 0 }^\infty  {\varphi \left( x \right){{\mathcal P}_1}\left( x \right){{\mathcal P}_0}\left( x \right)dx} }&{\int_{ 0 }^\infty  {\varphi \left( x \right){{\mathcal P}_1}^2\left( x \right)dx} }& \cdots &{\int_{ 0 }^\infty  {\varphi \left( x \right){{\mathcal P}_1}\left( x \right){{\mathcal P}_N}\left( x \right)dx} }\\
 \vdots & \vdots & \ddots & \vdots \\
{\int_{ 0 }^\infty  {\varphi \left( x \right){{\mathcal P}_N}\left( x \right){{\mathcal P}_0}\left( x \right)dx} }&{\int_{ 0 }^\infty  {\varphi \left( x \right){{\mathcal P}_N}\left( x \right){{\mathcal P}_1}\left( x \right)dx} }& \cdots &{\int_{ 0 }^\infty  {\varphi \left( x \right){{\mathcal P}_N}^2\left( x \right)dx} }
\end{array}} \right]},
\end{align}
\begin{align}\label{eqn:def_vec_b}
{\bf{b}} = {\left[ {\begin{array}{*{20}{c}}
{\int_0^\infty  {{f_{I_K^{IR}}}(x){{\cal P}_0}\left( x \right)dx} }&{\int_0^\infty  {{f_{I_K^{IR}}}(x){{\cal P}_1}\left( x \right)dx} }& \cdots &{\int_0^\infty  {{f_{I_K^{IR}}}(x){{\cal P}_N}\left( x \right)dx} }
\end{array}} \right]^{\rm{T}}},
\end{align}
\begin{align}\label{eqn:def_vec_d}
{\bf{d}} = {\left[ {\begin{array}{*{20}{c}}
{\int_0^\infty  {\varphi (x){\mathcal P_0}\left( x \right)dx} }&{\int_0^\infty  {\varphi (x){\mathcal P_1}\left( x \right)dx} }& \cdots &{\int_0^\infty  {\varphi (x){\mathcal P_N}\left( x \right)dx} }
\end{array}} \right]^{\rm{T}}},
\end{align}
\end{figure*}
and $c$ is
\begin{equation}\label{eqn:def_cons_c}
c = \int_{ 0 }^\infty  {\varphi \left( x \right){\psi ^2}\left( x \right)dx}.
\end{equation}

According to the Karush-Kuhn-Tucker (KKT) conditions, the optimal solutions ${\boldsymbol{\eta}} $ and $\varsigma $ should satisfy the following conditions
\begin{equation}\label{eqn:fir_der}
\left\{ {\begin{array}{*{20}{l}}
{\frac{{\partial \Lambda \left( {{\boldsymbol{\eta}} ,\varsigma |d\mu (x),N} \right)}}{{\partial {{\boldsymbol{\eta}} }}}{\rm{ = 2}}{\bf{A{\boldsymbol{\eta}}  }} - 2{\bf{b}} + \varsigma {\bf{d}} = 0}\\
{\frac{{\partial \Lambda \left( {{\boldsymbol{\eta}} ,\varsigma |d\mu (x),N} \right)}}{{\partial {\varsigma }}} {= {\boldsymbol{\eta}} ^{\rm{T}}}{{\bf{d}}-1} = 0}
\end{array}} \right..
\end{equation}
As proved in Appendix \ref{app:proof_inver_A}, the matrix $\bf A$ is invertible. Then the solution to (\ref{eqn:fir_der}) is unique and follows as
\begin{equation}\label{eqn:sol_to_fir_der}
\left\{ {\begin{array}{*{20}{l}}
{{\boldsymbol{\eta}} = {{\bf{A}}^{ - 1}}\left( {{\bf{b}} - \frac{\varsigma }{2}{\bf{d}}} \right)}\\
{\varsigma  = \frac{{2\left( {{{\bf{b}}^{\rm{T}}}{{\bf{A}}^{ - 1}}{\bf{d}} - 1} \right)}}{{{{\bf{d}}^{\rm{T}}}{{\bf{A}}^{ - 1}}{\bf{d}}}}}
\end{array}} \right.
\end{equation}

Clearly from (\ref{eqn:def_mtr_A}), (\ref{eqn:def_vec_b}), (\ref{eqn:def_vec_d}) and (\ref{eqn:sol_to_fir_der}), the monic orthogonal polynomials ${{\mathcal P}_n}\left( x \right)$ need to be determined before the calculation of the coefficient ${\boldsymbol{\eta}}$. In other words, the measure $d\mu (x)$ should be determined first. In what follows, the selection of the measure and the orthogonal polynomials and the calculation of the coefficient ${\boldsymbol{\eta}}$ are then discussed in detail.

\subsubsection{Selection of $d\mu (x)$ and $\boldsymbol{\mathcal{P}}(x)$}
After analyzing the MSE of the fitting error ${\rm{E}}\{ {e\left( x \right)^2}\}$, we have the following property.
\begin{property}\label{pro:irr}
For any measure $d\mu(x)$, the same minimal MSE can be attained. Moreover, the optimal polynomial $\hat \psi_N(x)$ is unique irrespective of the choice of $d\mu(x)$.
\end{property}
\begin{IEEEproof}
Please see Appendix \ref{app:proof_irre_to_pn}.
\end{IEEEproof}

Although the choice of the measure $d\mu (x)$ would not affect the solution of the optimal polynomial $\hat \psi_N(x)$ as shown in \textbf{Property 1}, it does affect the computational complexity in deriving the optimal polynomial. More specifically, the major computation comes from the inverse operation of the matrix $\bf A$ as seen from (\ref{eqn:sol_to_fir_der}). It is clear from the definition of $\bf A$ (\ref{eqn:def_mtr_A}) that the complexity of the inverse operation depends on the choice of the orthogonal polynomials ${{\mathcal P}_n}\left( x \right)$ and thus depends on the measure $d\mu (x)$. It is generally known that the complexity in computing the inverse of a $N \times N$ matrix could be significantly reduced from $\mathcal{O}(N^3)$ to $\mathcal{O}(N)$ when the matrix is diagonal. Therefore, to reduce the computational complexity, the measure $d\mu (x)$ is suggested to be chosen such that the matrix $\bf A$ is diagonal. Clearly from the structure of the matrix $\bf A$ (\ref{eqn:def_mtr_A}), when the measure is chosen as $d\mu (x) = {\varphi \left( x \right)}dx$, with the orthogonality of the polynomials in (\ref{eqn_dirac_def}), the matrix $\bf A$ will reduce to a diagonal matrix as
 \begin{equation}\label{eqn:diag_matr}
{\bf{A}} = \rm{diag}\left( {{\mathcal D_0},{\mathcal D_1}, \cdots ,{{\mathcal D}_N}} \right).
\end{equation}

Now with the measure $d\mu (x) = {\varphi \left( x \right)}dx$, the monic orthogonal polynomials ${{\mathcal P}_n}\left( x \right) = \sum\nolimits_{k = 0}^n {{C_{n,k}}{x^k}}$ with $C_{n,n}=1$ can be uniquely determined by the method introduced in \cite{gautschi2004orthogonal}. Specifically, the polynomial coefficients $C_{n,k}$ can be determined as follows. Following the three-term recurrence relation (TTRR) in \cite[Theorem 1.27]{gautschi2004orthogonal}, the monic orthogonal polynomials should satisfy
\begin{align}\label{eqn_orth_coeff}
&{{{\mathcal P}_{n + 1}}\left( x \right) = \left( {x - {\alpha _n}} \right){{\mathcal P}_n}\left( x \right) - {\beta _n}{{\mathcal P}_{n - 1}}\left( x \right),\, n = 0,1,2, \cdots ,} \notag\\
&{{{\mathcal P}_{ - 1}}\left( x \right) = 0,\;\;\;{\kern 1pt} {{\mathcal P}_0}\left( x \right) = 1,}
\end{align}
where
\begin{align}\label{eqn_expli_alpha}
{\alpha _n} &= \frac{\left<x{\mathcal P}_n(x),{\mathcal P}_n(x)\right>}{\left<{\mathcal P}_n(x),{\mathcal P}_n(x)\right>} \notag \\
&= \frac{{\sum\limits_{i = 0}^n {\sum\limits_{j = 0}^n {{C_{n,i}}{C_{n,j}}{\nu _{i + j + 1}}} } }}{{\sum\limits_{i = 0}^n {\sum\limits_{j = 0}^n {{C_{n,i}}{C_{n,j}}{\nu _{i + j}}} } }},\,n = 0,1,2,\cdots,
\end{align}
\begin{align}\label{eqn_expli_beta}
{\beta _n} &= \frac{\left<{\mathcal P}_n(x),{\mathcal P}_n(x)\right>}{\left<{\mathcal P}_{n-1}(x),{\mathcal P}_{n-1}(x)\right>} \notag \\
 &=\frac{{\sum\limits_{i = 0}^n {\sum\limits_{j = 0}^n {{C_{n,i}}{C_{n,j}}{\nu _{i + j}}} } }}{{\sum\limits_{i = 0}^{n - 1} {\sum\limits_{j = 0}^{n - 1} {{C_{n - 1,i}}{C_{n - 1,j}}{\nu _{i + j}}} } }},\,n = 1,2,\cdots.
\end{align}
and $\nu _n$ denotes the $n$th moment with respect to the cumulative distribution function (CDF) $\mu (x)$, that is,
\begin{align}\label{eqn:deter_mom}
{\nu _n} &= \int_0^\infty  {{x^n}d\mu \left( x \right)}  = \int_0^\infty  {{x^n}\varphi \left( x \right)dx}  \notag\\
 &= \int_0^\infty  {\frac{{{x^{n + \zeta  - 1}}{e^{ - \frac{x}{\theta }}}}}{{{\theta ^\zeta }\Gamma \left( \zeta  \right)}}dx} = \frac{{{\theta ^n}\Gamma \left( {n + \zeta } \right)}}{{\Gamma \left( \zeta  \right)}}
\end{align}
With ${{\mathcal P}_n}\left( x \right) = \sum\nolimits_{k = 0}^n {{C_{n,k}}{x^k}}$ and the relation in (\ref{eqn_orth_coeff}), the polynomial coefficients $C_{n,k}$ can be obtained recursively as
\begin{equation}\label{eqn_c_n_k}
{C_{n + 1,k}} = \left\{ {\begin{array}{*{20}{c}}
{{C_{n,k - 1}} - {\alpha _n}{C_{n,k}} - {\beta _n}{C_{n - 1,k}},}&{ 0 \le k \le n + 1;}\\
{0,}&{else.}
\end{array}} \right.
\end{equation}
with ${C_{0,0}} = 1$. Then the parameter ${\mathcal D}_n$ in (\ref{eqn_dirac_def}) can be determined as
\begin{equation}\label{eqn:deter_D}
{{\mathcal D}_n} = \sum\limits_{i = 0}^n {\sum\limits_{j = 0}^n {{C_{n,i}}{C_{n,j}}{\nu _{i + j}}} }.
\end{equation}
Specifically, ${{\mathcal D}_0}=1$.

\subsubsection{Calculation of $\boldsymbol \eta$}
To compute $\boldsymbol \eta$ in (\ref{eqn:sol_to_fir_der}), the vectors $\bf b$, $\bf d$ and $\varsigma $ should be determined first. By  the definition of $\mathcal M(i)$ and integrating out $x$ for (\ref{eqn:def_vec_b}), the vector $\bf{b}$ can be consequently written as
\begin{multline}\label{eqn:vec_b_rew1}
{\bf{b}} = \left[ {\begin{array}{*{20}{c}}
{\sum\limits_{k = 0}^0 {{C_{0,k}}} {\cal M}\left( k \right)}&{\sum\limits_{k = 0}^1 {{C_{0,k}}} {\cal M}\left( k \right)}
\end{array}} \right.\\
{\left. {\begin{array}{*{20}{c}}
 \cdots &{\sum\limits_{k = 0}^N {{C_{N,k}}} {\cal M}\left( k \right)}
\end{array}} \right]^{\rm{T}}}.
\end{multline}

Considering $\mathcal {P}_0(x) = 1$ and the orthogonality among the polynomials $\mathcal {P}_n(x)$, the vector {\bf{d}} given in (\ref{eqn:def_vec_d}) is directly reduced as
\begin{equation}\label{eqn:d_exp_det}
{\bf{d}} = {\left[ {1,\quad \overbrace {0,\quad  \cdots \quad ,0}^{N - terms}} \right]^{\rm T}}.
\end{equation}

On the other hand, by putting (\ref{eqn:diag_matr}), (\ref{eqn:vec_b_rew1}) and (\ref{eqn:d_exp_det}) into (\ref{eqn:sol_to_fir_der}), the Lagrange multiplier $\varsigma $ is calculated as
\begin{equation}\label{eqn:det_varsigma}
\varsigma  = \frac{{{C_{0,0}}{\mathcal M}\left( 0 \right){D_0} - 1}}{{{D_0}}} = 0
\end{equation}
by recalling that $C_{0,0}=1$, ${\mathcal M}(0)=1$ and $\mathcal D_0=1$. Accordingly, the coefficients $\boldsymbol \eta$ can be finally computed as
\begin{multline}\label{eqn:eta_exp}
{\boldsymbol{\eta }} = {{\bf{A}}^{ - 1}}{\bf{b}} \\
= {\left[ {\begin{array}{*{20}{c}}
{\frac{{\sum\limits_{k = 0}^0 {{C_{0,k}}} {\mathcal M}\left( k \right)}}{{{{\mathcal D}_0}}}}&{\frac{{\sum\limits_{k = 0}^1 {{C_{1,k}}} {\mathcal M}\left( k \right)}}{{{{\mathcal D}_1}}}}& \cdots &{\frac{{\sum\limits_{k = 0}^N {{C_{N,k}}} {\mathcal M}\left( k \right)}}{{{{\mathcal D}_N}}}}
\end{array}} \right]^{\rm T}}.
\end{multline}
\subsection{Discussions}
With the approximated PDF of the accumulated mutual information in (\ref{eqn:reulst_PDF}), the outage probability which is equivalent to the CDF of the accumulated mutual information can be easily obtained. From the expression of the outage probability, some interesting insights could be found in the following. Moreover, as shown in (\ref{eqn_op}), the MSE of the fitting error also depends on the degree of the polynomials $N$. The selection of the degree will also be briefly discussed here.

\subsubsection{Insights of Outage Probability}
After determining $\boldsymbol \eta$, the approximated PDF $f_{{{\boldsymbol \gamma_{1:K}}},N}^{IR}(x)$ is expressed with the expansion of ${{{\mathcal P}_n}\left( x \right)}$ as
\begin{equation}\label{eqn:appro_pdf_exp}
f_{{{\boldsymbol \gamma_{1:K}}},N}^{IR}(x) = \varphi (x)\sum\limits_{n = 0}^N {{\eta _n}\sum\limits_{i = 0}^n {{C_{n,i}}{x^i}} }.
\end{equation}

Henceforth, the approximated CDF for $I_K^{IR}$ can be obtained as
\begin{align}\label{eqn_exp_cdf}
{F_{I_K^{IR}}}(x) &\approx F_{{{\boldsymbol \gamma_{1:K}}},N}^{IR}(x) = \int_0^x {f_{{{\boldsymbol \gamma_{1:K}}},N}^{IR}(t)dt} \notag\\
& = \sum\limits_{n = 0}^N {{\eta _n}\sum\limits_{i = 0}^n {{C_{n,i}}\int_0^x {{t^i}\varphi (t)dt} } }.
\end{align}

To facilitate the analysis, a family of functions $W_i(x)$ is defined as
\begin{align}\label{eqn_W_i}
{W_i}\left( x \right) &= \frac{1}{{{\nu _i}}}\int_0^x {{t^i}\varphi \left( t \right)dt}  = \frac{1}{{{\theta ^{i + \zeta }}\Gamma \left( {i + \zeta } \right)}}\int_0^x {{t^{i + \zeta  - 1}}{e^{ - \frac{t}{\theta }}}dt} \notag\\
 & = \frac{{\gamma \left( {i + \zeta ,\frac{x}{\theta }} \right)}}{{\Gamma \left( {i + \zeta } \right)}}
\end{align}
where $\gamma(\cdot)$ represents the lower incomplete Gamma function. It is clear that $W_i(x)$ is the CDF of a Gamma RV with parameters $(i+\zeta,\theta)$. Consequently, by exchanging the order of summations, the CDF in (\ref{eqn_exp_cdf}) can be rewritten as
\begin{align}\label{eqn_CDF_final}
F_{{{\boldsymbol \gamma_{1:K}}},N}^{IR}(x) &= \sum\limits_{n = 0}^N {{\eta _n}\sum\limits_{i = 0}^n {{C_{n,i}}{\nu _i}{W_i}\left( x \right)} } \notag \\
&= \sum\limits_{i = 0}^N {{W_i}\left( x \right){\nu _i}\sum\limits_{n = i}^N {{\eta _n}} {C_{n,i}}}  \notag \\
&= \sum\limits_{i = 0}^N {{\kappa _i}} {W_i}\left( x \right)
\end{align}
where
\begin{equation}\label{eqn_k_i}
{\kappa _i} = {\nu _i}\sum\limits_{n = i}^N {{\eta _n}} {C_{n,i}}.
\end{equation}
Meanwhile, the parameters $\{\kappa_i\}$ satisfy
\begin{multline}\label{eqn:sum_kappa}
\mathop {\lim }\limits_{x \to \infty } F_{{{\boldsymbol \gamma_{1:K}}},N}^{IR}(x) = \mathop {\lim }\limits_{x \to \infty } \sum\limits_{i = 0}^N {{\kappa _i}} {W_i}\left( x \right) = 1 \\
\Rightarrow \sum\limits_{i = 0}^N {{\kappa _i}}  = 1.
\end{multline}

Therefore, under time-correlated Rayleigh fading channels, the CDF of the accumulated mutual information $F_{I_K^{IR}}(x)$ can be written as the weighted sum of the CDFs of Gamma RVs, such that
 \begin{equation}\label{eqn:appro_acmu}
{F_{I_K^{IR}}}(x) \approx \sum\limits_{i = 0}^N {{\kappa _i}} {W_i}\left( x \right).
 \end{equation}

It means that the outage probability $P_{out}^{IR}\left( K \right) = F_{I_K^{IR}}({\cal R})$ can be written as the weighted sum of a number of outage probabilities, each associated with one Gamma RV\footnote{The source code of our approximation is available at \\ {\it{\myself}}.}. Specifically, if $N=0$, the approximation reduces to the ordinary Gamma approximation which is valid for the case with \textit{independent} fading channels \cite{Chelli2014performance}. The approximation in (\ref{eqn:appro_acmu}) can ease the analysis of the system behaviors with respect to various system parameters and facilitate system design to achieve various objectives, e.g., the optimal rate design to maximize the long term average throughput. This will be further illustrated in Section V.
\subsubsection{Choice of $N$}
As shown in  \textbf{Property \ref{pro:irr}}, the same minimal MSE ${{\cal S}_{mse}}({\boldsymbol{\eta }}|d\mu (x),N)$ can be attained whatever the measure $d\mu(x)$ is.  Hereby, we define ${{\mathcal S}_{min\_mse}}({\boldsymbol{\eta }}|N)$ as the minimal MSE given $N$. By substituting (\ref{eqn:sol_to_fir_der}) into the objective function of MSE in (\ref{eqn_op}), it yields the minimal MSE as
\begin{align}\label{eqn:mse_det}
{{\mathcal S}_{min\_mse}}({\boldsymbol{\eta }}|N) &= {{\boldsymbol{\eta }}^{\rm{T}}}\left( {{\bf{A}\boldsymbol{\eta} } - 2{\bf{b}}} \right) + c \notag \\
 &= c - {{\boldsymbol{\eta }}^{\rm{T}}}\left( {{\bf{b}} + \frac{\varsigma }{2}{\bf{d}}} \right) \notag \\
 &= c - \left( {{{\bf{b}}^{\rm T}} - \frac{\varsigma }{2}{{\bf{d}}^{\rm{T}}}} \right){{\bf{A}}^{ - 1}}\left( {{\bf{b}} + \frac{\varsigma }{2}{\bf{d}}} \right).
\end{align}

\begin{figure}
  \centering
  \includegraphics[width=3.5in]{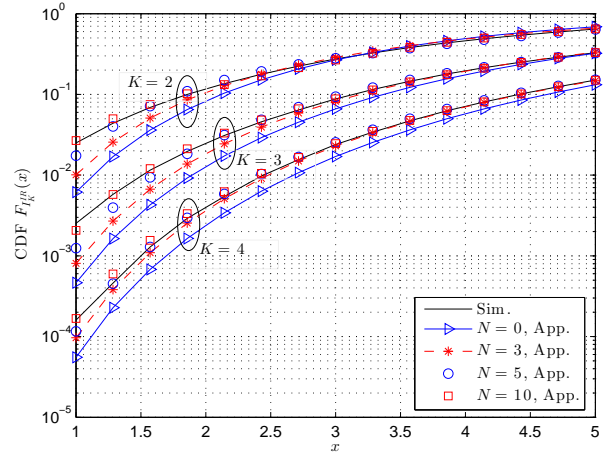}\\
  \caption{Comparison between the approximated CDFs ${F_{I_K^{IR}}}(x)$  with different $N$ and the true CDFs obtained from Monte-Carlo simulations.}\label{fig:conver}
\end{figure}

Putting (\ref{eqn:diag_matr}), (\ref{eqn:vec_b_rew1}), (\ref{eqn:d_exp_det}) and (\ref{eqn:det_varsigma}) into (\ref{eqn:mse_det}), the minimal MSE is rewritten as
\begin{multline}\label{eqn:MSE_EXP}
{{\mathcal S}_{min\_mse}}({\boldsymbol{\eta }}|N) = \int_0^\infty  {\varphi \left( x \right){\psi ^2}\left( x \right)dx}\\
 - \sum\limits_{n = 0}^N {{{\mathcal D}_n}^{ - 1}{{\left( {\sum\limits_{k = 0}^n {{C_{n,k}}} {\mathcal M}\left( k \right)} \right)}^2}} \ge 0.
\end{multline}
Clearly, the minimum MSE decreases as the degree $N$ increases, i.e. ${{\mathcal S}_{min\_mse}}({\boldsymbol{\eta }}|N) \ge {{\mathcal S}_{min\_mse}}({\boldsymbol{\eta }}|N+1)$, since ${\mathcal D_n} = \left\langle {{{\mathcal P}_n}\left( x \right),{{\mathcal P}_n}\left( x \right)} \right\rangle  > 0$. It implies that the accuracy of the PDF approximation is limited by the degree $N$ and would be improved as the degree $N$ increases. This result can be further demonstrated by Fig. \ref{fig:conver} where the approximated CDFs of $I_K^{IR}$ using different degrees are compared with the true CDFs obtained from Monte-Carlo simulations, by taking a system with the following setting as an example: the channel correlation coefficient $\rho_{k,l}=0.5$ and the mean of the SNR ${\rm{E}}(\gamma_l)=2{\sigma _l'}^2=5$, where $1 \le k \ne l \le K$.

However, as shown in Fig. \ref{fig:conver}, the improvement of the approximation accuracy becomes minor when the degree becomes relatively large. Moreover, the increase of the degree $N$ would also cause the increase of the computational complexity. It is thus necessary to properly choose the degree to balance the approximation accuracy and computational complexity. To quantify the contribution of increasing the degree in terms of the approximation accuracy, a metric of MSE reduction is defined as
\begin{align}\label{eqn:impro_orth_polys}
{\Delta _N } &\triangleq {{\cal S}_{min\_mse}}({\boldsymbol{\eta }}|0) - {{\cal S}_{min\_mse}}({\boldsymbol{\eta }}|N)\notag \\
&=\sum\limits_{n = 1}^N {{{\mathcal D}_n}^{ - 1}{{\left( {\sum\limits_{k = 0}^n {{C_{n,k}}} {\mathcal M}\left( k \right)} \right)}^2}}.
\end{align}
For a good balance between approximation accuracy and computational complexity, the degree is suggested to be chosen as
\begin{equation}\label{eqn:N_app}
{N} = \min \left\{ {\min \left\{ {n\left| {{r_n} \le \epsilon } \right.} \right\},\hat N} \right\}
\end{equation}
where $\hat N$ is a pre-determined upper bound of the degree to limit the computational complexity, and $\epsilon$ denotes the tolerance for normalized MSE reduction defined as ${r_n} = \frac{{{\Delta _n - {\Delta _{n-1}}}}}{{{\Delta _n}}}$. Clearly, ${{r_N} \le \epsilon}$ holds. It roughly indicates that no significant improvement on MSE can be expected if the degree is larger than $N$ in (\ref{eqn:N_app}). Therefore the degree of $N$ in (\ref{eqn:N_app}) is sufficient for a good approximation. Notice that $\Delta _1 = {\Delta _2} = 0$ by using the orthogonality of the polynomials. Thus the upper bound $\hat N$ and the degree $N$ should be set greater than $2$ without doubt.
\section{Diversity Order}
\label{sec:dive}
Basically, HARQ-IR schemes exploit not only coding gain but also time diversity to improve the transmission reliability. To better understand the behavior of HARQ-IR schemes, diversity order which is another important performance metric is also analyzed in this paper. To facilitate the analysis, the transmit SNR in each HARQ round is set equal, i.e. $P_1/\mathfrak{N}_1 = P_2/\mathfrak{N}_2 = \cdots =P_M/\mathfrak{N}_M=\gamma_T$. The diversity order $d$ for the HARQ scheme is defined as \cite{Chelli2014performance,zheng2003diversity}
\begin{equation}\label{eqn:diver_def_org}
d =  - \mathop {\lim }\limits_{{\gamma _T} \to \infty } \frac{{\log \left( {P_e} \right)}}{{\log \left( {{\gamma _T}} \right)}},
\end{equation}
where $P_e$ denotes the error probability. As shown in \cite{Chelli2014performance,zheng2003diversity}, the error probability can be well approximated as the outage probability ${P_{out}^{IR}\left( M \right)}$ when a capacity achieving code is applied. Therefore the diversity order $d$ can be well approximated as
\begin{equation}\label{eqn:diver_def}
d =  - \mathop {\lim }\limits_{{\gamma _T} \to \infty } \frac{{\log \left( {P_{out}^{IR}\left( M \right)} \right)}}{{\log \left( {{\gamma _T}} \right)}}.
\end{equation}
Namely, diversity order quantifies the slope of $-\log \left( {P_{out}^{IR}\left( M \right)} \right)$ with respect to $\log \left( {{\gamma _T}} \right)$ when ${{\gamma _T} \to \infty }$.

Owing to the following inequalities
\begin{multline}\label{eqn:ineqs}
{\log _2}\left( {1 + \sum\limits_{l = 1}^M {\gamma_l} } \right) \le {\log _2}\left( {\prod\limits_{l = 1}^M {\left( {1 + \gamma_l} \right)} } \right) \\
 \le M{\log _2}\left( {1 + {M^{ - 1}}\sum\limits_{l = 1}^M {\gamma_l} } \right),
\end{multline}
the outage probability can be bounded as
\begin{multline}\label{eqn:prob_ineqs}
\Pr \left( {M{{\log }_2}\left( {1 + {M^{ - 1}}\sum\limits_{l = 1}^M {\gamma_l} } \right) < {\mathcal R}} \right) \le P_{out}^{IR}\left( M \right) \\
\le \Pr \left( {{{\log }_2}\left( {1 + \sum\limits_{l = 1}^M {\gamma_l} } \right) < {\mathcal R}} \right),
\end{multline}
where the right inequality in (\ref{eqn:ineqs}) holds by using the Jensen's inequality. Defining $Y = \sum\nolimits_{l = 1}^M {\gamma_l} $, the bounds in (\ref{eqn:prob_ineqs}) can be rewritten as
\begin{equation}\label{eqn:out_ineqs_rew}
\Pr \left( {Y < M( {{2^{{M^{ - 1}}{\cal R}}} - 1})} \right) \le P_{out}^{IR}\left( M \right) \le \Pr \left( {Y < {2^{\cal R}} - 1} \right).
\end{equation}
Clearly, $Y$ represents a sum of correlated exponential RVs $\{\gamma_l\}_{l=1}^M$. The CDF of a sum of correlated exponential RVs has been derived in \cite{kalyani2012asymptotic} and the result is summarized as the following theorem.

\begin{theorem}\cite{kalyani2012asymptotic}
\label{theorem:CDF}
Given the joint PDF regarding to exponential RVs $\{\gamma_l\}_{l=1}^M$ in (\ref{eqn_joint_pdf_X}), the CDF of $Y = \sum\nolimits_{l = 1}^M {\gamma_l} $ can be obtained as
\begin{multline}\label{eqn:cdf_Y_def_in_meijer_G}
{F_Y}\left( y \right) = \Pr \left( {Y < y} \right) = \frac{{{y^M}}}{{\det \left( {\bf{B}} \right) \Gamma \left( {M + 1} \right)}}\times\\
\Phi _2^{\left( M \right)}\left( {1, \cdots ,1;M + 1; - {\delta _1}^{ - 1}y, \cdots , - {\delta _M}^{ - 1}y} \right)
\end{multline}
where the notation $\rm det(\cdot)$ represents the determinant operation, $\Phi _2^{\left( M \right)}\left(  \cdot  \right)$ denotes the confluent Lauricella function \cite[Def. A.19]{mathai2009h}, $\{\delta _k\}_{k=1}^{M}$ are defined as the eigenvalues of the matrix $\bf B=FE$, where $\bf F$ is an $M \times M$ diagonal matrix with diagonal entries as $\{2{\sigma _k'}^2\}_{k=1}^{M}$, and $\bf E$ is an $M \times M$ positive definite matrix given by
\begin{equation}\label{eqn:C_mat_def}
{\bf{E}} = \left[ {\begin{array}{*{20}{c}}
1&{\sqrt {{\rho _{1,2}}} }& \cdots &{\sqrt {{\rho _{1,M}}} }\\
{\sqrt {{\rho _{2,1}}} }&1& \cdots &{\sqrt {{\rho _{2,M}}} }\\
 \vdots & \vdots & \ddots & \vdots \\
{\sqrt {{\rho _{M,1}}} }&{\sqrt {{\rho _{M,2}}} }& \cdots &1
\end{array}} \right], \, 0 \le \rho_{k,l} <1.
\end{equation}
\end{theorem}

Substituting (\ref{eqn:out_ineqs_rew}) into (\ref{eqn:diver_def}), the bounds of the diversity order can be found as
\begin{multline}\label{eqn:diversity_bet1}
 - \mathop {\lim }\limits_{{\gamma _T} \to \infty } \frac{{\log \left( {{F_Y}\left( {{2^{\mathcal R}} - 1} \right)} \right)}}{{\log \left( {{\gamma _T}} \right)}} \le d \\
 \le  - \mathop {\lim }\limits_{{\gamma _T} \to \infty } \frac{{\log \left( {{F_Y}\left( {M\left( {{2^{{M^{ - 1}}{\cal R}}} - 1} \right)} \right)} \right)}}{{\log \left( {{\gamma _T}} \right)}}.
\end{multline}

By using (\ref{eqn:cdf_Y_def_in_meijer_G}), the left inequality in (\ref{eqn:diversity_bet1}) can be rewritten as
\begin{align}\label{eqn:limit_ineq}
&d \ge \mathop {\lim }\limits_{{\gamma _T} \to \infty } \left[ {\frac{{\log \left( {\det \left( {\bf{B}} \right)} \right)}}{{\log \left( {{\gamma _T}} \right)}}} \right. - \notag \\
&{\left. {\frac{{\log \left( {\Phi _2^{\left( M \right)}\left( {1, \cdots ,1;M + 1; - {\delta _1}^{ - 1}y, \cdots , - {\delta _M}^{ - 1}y} \right)} \right)}}{{\log \left( {{\gamma _T}} \right)}}} \right]_{y = {2^R} - 1}}.
\end{align}
The first term on the right hand side of (\ref{eqn:limit_ineq}) can be simplified by using the property of determinants as
\begin{equation}\label{eqn:deter_simple}
\mathop {\lim }\limits_{{\gamma _T} \to \infty } \frac{{\log \left( {\det \left( {\bf{B}} \right)} \right)}}{{\log \left( {{\gamma _T}} \right)}} = \mathop {\lim }\limits_{{\gamma _T} \to \infty } \frac{{\log \left( {\det \left( {\bf{F}} \right)} \right)}}{{\log \left( {{\gamma _T}} \right)}} + \mathop {\lim }\limits_{{\gamma _T} \to \infty } \frac{{\log \left( {\det \left( {\bf{E}} \right)} \right)}}{{\log \left( {{\gamma _T}} \right)}}.
\end{equation}
Recalling that ${\bf{F}} = {\gamma _T}{\rm {diag}}\left( {2{\sigma _1}^2, \cdots ,2{\sigma _M}^2} \right)$ and ${\bf{E}}$ is irrelevant with ${\gamma _T}$, (\ref{eqn:deter_simple}) can be reduced to
\begin{equation}\label{eqn:deter_simpl_lim}
\mathop {\lim }\limits_{{\gamma _T} \to \infty } \frac{{\log \left( {\det \left( {\bf{B}} \right)} \right)}}{{\log \left( {{\gamma _T}} \right)}} = M.
\end{equation}
To derive the second term on the right hand side of the inequality (\ref{eqn:limit_ineq}), we define ${\bf G} = {\rm{diag}}\left( {2{\sigma _1}^2, \cdots ,2{\sigma _M}^2} \right)\bf{E}$ for notational simplicity, such that ${\bf{B}}=\gamma_T \bf{G}$.
By using the series representation of the confluent Lauricella function \cite{aalo2005another}, the confluent Lauricella function in the second term on the right hand side of the inequality (\ref{eqn:limit_ineq}) can be rewritten as
\begin{multline}\label{eqn:limit}
\Phi _2^{\left( M \right)}\left( {1, \cdots ,1;M + 1; - {\delta _1}^{ - 1}y, \cdots , - {\delta _M}^{ - 1}y} \right) \\
 = 1 + \sum_{{m_1+\cdots+m_M} > 0} {  {\frac{{\prod\limits_{k = 1}^M {{{\left( { - {\delta _k}^{ - 1}y} \right)}^{{m_k}}}} }}{{{{\left( M+1 \right)}_{{m_1} +  \cdots  + {m_M}}}}}} } ,
\end{multline}
where $(\cdot)_n$ denotes Pochhammer symbol. Since
\begin{align}\label{eqn:der_f_inf_gamma_conf}
&\left | \sum_{{m_1+\cdots+m_M} > 0} {  {\frac{{\prod\limits_{k = 1}^M {{{\left( { - {\delta _k}^{ - 1}y} \right)}^{{m_k}}}} }}{{{{\left( M+1 \right)}_{{m_1} +  \cdots  + {m_M}}}}}} } \right | \notag\\
&\leq \sum_{{m_1+\cdots+m_M} > 0} {{\frac{{{m_1}! \cdots {m_M}!}}{{{{\left( {M + 1} \right)}_{{m_1} +  \cdots  + {m_M}}}}}\frac{{\prod\limits_{k = 1}^M  {{{\left| { - {\delta _k}^{ - 1}y} \right|}^{{m_k}}}} }}{{{m_1}! \cdots {m_M}!}}} } \notag \\
& \underset{(a)}{\leq } \sum_{{m_1+\cdots+m_M} > 0} {\frac{{\prod\limits_{k = 1}^M {{{ \left | { - {\delta _k}^{ - 1}y} \right | }^{{m_k}}}} }}{{{m_1}! \cdots {m_M}!}}}  \nonumber \\
& = \sum\limits_{L = 1}^\infty  {\frac{1}{{L!}}\sum_{\sum\limits_{k = 1}^M {{m_k} = L} } {\frac{{L!}}{{{m_1}! \cdots {m_M}!}}\prod\limits_{k = 1}^M {{{\left| { - {\delta _k}^{ - 1}y} \right|}^{{m_k}}}} } } \nonumber \\
& = \sum\limits_{L = 1}^\infty  {\frac{1}{{L!}}{{\left( {  y\sum\limits_{k = 1}^M {\left |{\delta _k} \right |^{ - 1}} } \right)}^L}} \notag \\
& = {e^{  y\sum\limits_{k = 1}^M {\left| {\delta _k} \right|^{ - 1}} }} -1 \underset{(b)} {=} {e^{  y\sum\limits_{k = 1}^M {\left| {\gamma_T \beta _k} \right|^{ - 1}} }} -1 \notag \\
&= {e^{  y  \gamma_T^{-1}\sum\limits_{k = 1}^M {\left| {\beta _k} \right|^{ - 1}} }} -1
\end{align}
where $(a)$ follows from ${m_1}! \cdots {m_M}! \le {\left( {M + 1} \right)_{{m_1} +  \cdots  + {m_M}}}$, $\beta_k$ denotes the eigenvalues of $\bf G$ and $(b)$ comes from ${\bf{B}}=\gamma_T \bf{G}$, together with $\mathop {\lim }\limits_{{\gamma _T} \to \infty } {e^{  y  \gamma_T^{-1}\sum\nolimits_{k = 1}^M {\left| {\beta _k} \right|^{ - 1}} }} =1$, we have
\begin{equation}\label{eqn:sec_term_limi_1}
\mathop {\lim }\limits_{{\gamma _T} \to \infty } \sum_{{m_1+m_2+\cdots+m_M} > 0} {  {\frac{{\prod\limits_{k = 1}^M {{{\left( { - {\delta _k}^{ - 1}y} \right)}^{{m_k}}}} }}{{{{\left( M+1 \right)}_{{m_1} +  \cdots  + {m_M}}}}}} }  = 0.
\end{equation}
It follows the limit of the confluent Lauricella function as
\begin{equation}\label{eqn:inf_gamm_conf}
\mathop {\lim }\limits_{{\gamma _T} \to \infty } \Phi _2^{\left( M \right)}\left( {1, \cdots ,1;M + 1; - {\delta _1}^{ - 1}y, \cdots , - {\delta _M}^{ - 1}y} \right) = 1.
\end{equation}
and then the second term on the right hand side of the inequality (\ref{eqn:limit_ineq}) reduces as
\begin{multline}\label{eqn:sec_term_limi}
\mathop {\lim }\limits_{{\gamma _T} \to \infty } \frac{{\log \left( {\Phi _2^{\left( M \right)}\left( {1, \cdots ,1;M + 1; - {\delta _1}^{ - 1}y, \cdots , - {\delta _M}^{ - 1}y} \right)} \right)}}{{\log \left( {{\gamma _T}} \right)}} \\
= 0.
\end{multline}

Substituting (\ref{eqn:deter_simpl_lim}) and (\ref{eqn:sec_term_limi}) into (\ref{eqn:limit_ineq}) leads to $d \ge M$. Following the same approach, the second inequality of (\ref{eqn:diversity_bet1}) can also be derived as $d \le M$. As a result, under time correlated fading channels with $0 \le \rho_{k,l} < 1$, the diversity order $d$ is equal to the number of transmissions $M$, i.e., $d=M$. Equivalently,
\begin{equation}\label{eqn:large_SNR_out}
{P_{out}^{IR}\left( M \right)} \propto  \frac{1}{{\gamma_T}^M}
\end{equation}
for large SNR. More precisely, the outage probability ${P_{out}^{IR}\left( M \right)}$ can be expressed as \cite[3.158]{tse2005fundamentals}
\begin{equation}
\label{eqn:outage}
P_{out}^{IR}\left( M \right) = c \left( {{\gamma _T},\sigma_k,{\rho_{k,l}}},\mathcal R,M \right) {{{\gamma _T}^{-M}}},
\end{equation}
where the coefficient $c\left( {{\gamma _T},{\sigma _k},{\rho_{k,l}},{\cal R},M} \right)$ satisfies
\begin{equation}\label{eqn:satis_c}
\mathop {\lim }\limits_{{\gamma _T} \to \infty } \frac{{\log \left( {c \left( {{\gamma _T},\sigma_k,{\rho_{k,l}},{\cal R},M} \right)} \right)}}{{\log \left( {{\gamma _T}} \right)}} = 0.
\end{equation}
From (\ref{eqn:outage}) and (\ref{eqn:satis_c}), we can find that when $\gamma _T$ is large, the slope of $-\log \left( {P_{out}^{IR}\left( M \right)} \right)$ with respect to $\log \left( {{\gamma _T}} \right)$ tends to be constant as $M$. In other words, when $\rho_{k,l} <1$, the time correlation ${\rho_{k,l}}$ would not affect the diversity order and the diversity order is constant as $M$. However the time correlation would influence the coefficient $c\left( {{\gamma _T},{\sigma _k},{\rho_{k,l}},{\cal R},M} \right)$ and thus affect the outage performance. This effect will be further investigated in Section V.

The above analysis indicates that full diversity can be achieved by HARQ-IR schemes even under time-correlated fading channels with $0 \le \rho_{k,l} <1$, which further justifies the benefit of HARQ-IR.

\textit{\textbf{Remark 1}}: The result of the diversity order is not applicable to the case with fully correlated fading channels. Under fully correlated Rayleigh fading channels, $|h_1| = |h_2| =\cdots=|h_M| \sim Rayleigh(2 {\sigma }^2)$ and $\rho_{k,l}=1$. The outage probability $P_{out}^{IR}\left( M \right)$ can be easily derived as
\begin{equation}\label{eqn:rho_1_outage}
P_{out}^{IR}\left( M \right) = 1 - \exp \left( { - \frac{{{2^{\mathcal R/M}} - 1}}{{{2 \sigma ^2 \gamma _T}}}} \right).
\end{equation}
Putting (\ref{eqn:rho_1_outage}) into (\ref{eqn:diver_def}), and by applying L'H\^opital's rule and the method of replacement with equivalent infinitesimal, the diversity order follows as
\begin{align}\label{eqn:rho_1_diversity_order}
d &=  - \mathop {\lim }\limits_{{\gamma _T} \to \infty } \frac{{\log \left( {1 - \exp \left( { - \frac{{{2^{{\cal R}/M}} - 1}}{{{2 \sigma ^2 \gamma _T}}}} \right)} \right)}}{{\log \left( {{\gamma _T}} \right)}} \notag \\
& = \mathop {\lim }\limits_{{\gamma _T} \to \infty } \frac{{\left( {{2^{{\cal R}/M}} - 1} \right)\exp \left( { - \frac{{{2^{{\cal R}/M}} - 1}}{{{2 \sigma ^2 \gamma _T}}}} \right)}}{{{2 \sigma ^2 \gamma _T}\left( {1 - \exp \left( { - \frac{{{2^{{\cal R}/M}} - 1}}{{{2 \sigma ^2 \gamma _T}}}} \right)} \right)}} \notag\\
 &= \mathop {\lim }\limits_{{\gamma _T} \to \infty } \frac{{{2^{{\cal R}/M}} - 1}}{{{2^{{\cal R}/M}} - 1}} = 1.
\end{align}
It means that under fully correlated fading channels, the diversity order is reduced to one.
\section{Numerical Results and Discussions}
\label{sec_sim}
With the above analytical results, the performance of HARQ-IR over time-correlated fading channels can be evaluated and optimal design of transmission scheme is also enabled. In the following, we take systems with ${\rm{E}}(|h_l|^2)= 2\sigma_l^2 =1$ and $\rho_{k,l}=\rho$ for $1 \le k \ne l \le M$ as examples for performance evaluation and optimal design. 

\subsection{Verification of Analytical Results}
\begin{figure}
  \centering
  \includegraphics[width=3.5in]{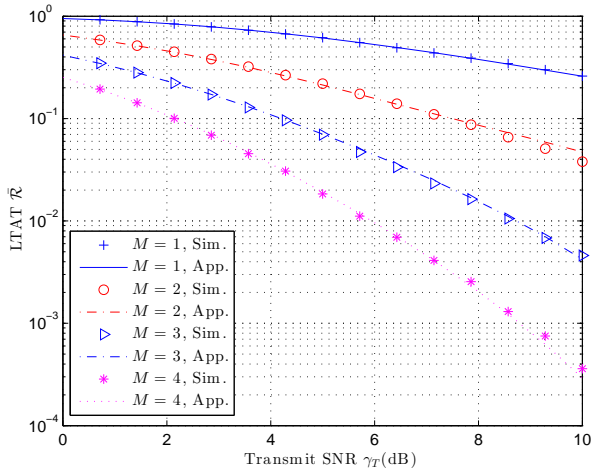}\\
  \caption{Outage Probability $P_{out}^{IR}(M)$ versus transmit SNR $\gamma_T$.}\label{fig_outage_probability}
\end{figure}

To verify our analytical expressions for the performance metrics of HARQ-IR over time-correlated Rayleigh fading channels, Monte-Carlo simulations are conducted for comparison. For illustration, we set $\mathcal R = 2 \rm{bps/Hz}$ and $\rho=0.5$, and the outage probability versus transmit SNR $\gamma_T$ is shown in Fig. \ref{fig_outage_probability}. Apparently, there is a perfect match between the analytical results and simulation results, which demonstrates the correctness of our analytical results. Moreover, as expected, the outage probability $P_{out}^{IR}(M)$ decreases as $M$ increases. For example, the outage probability for $M=1$ is about $2.6*10^{-1}$ for a transmit SNR of $10~\rm{dB}$. When $M$ is increased to $4$, the outage probability significantly drops to $3*10^{-4}$. It demonstrates a notable performance gain of HARQ-IR schemes.

\subsection{Impact of Time Correlation}
\begin{figure}
  \centering
  \includegraphics[width=3.5in]{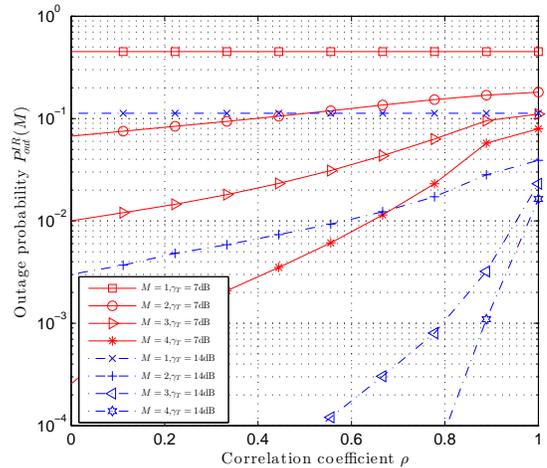}\\
  \caption{Outage Probability $P_{out}^{IR}\left( M \right)$ versus correlation coefficient $\rho$.}\label{fig_out_pro_rho}
\end{figure}

\begin{figure}
  \centering
  \includegraphics[width=3.5in]{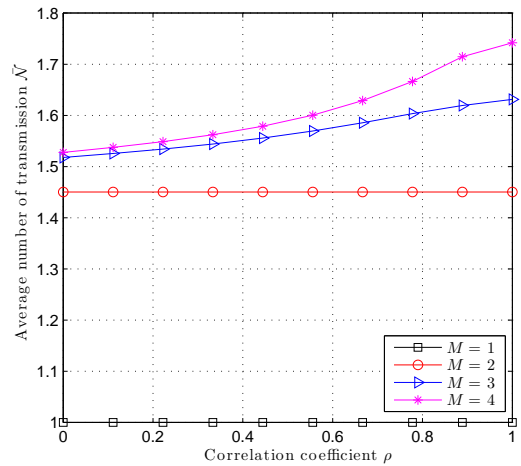}\\
  \caption{Average number of transmissions $\bar {\mathcal N}$ versus correlation coefficient $\rho$. ($\gamma_T=7$dB)}\label{fig_avg_num_rho}
\end{figure}

\begin{figure}
  \centering
  \includegraphics[width=3.5in]{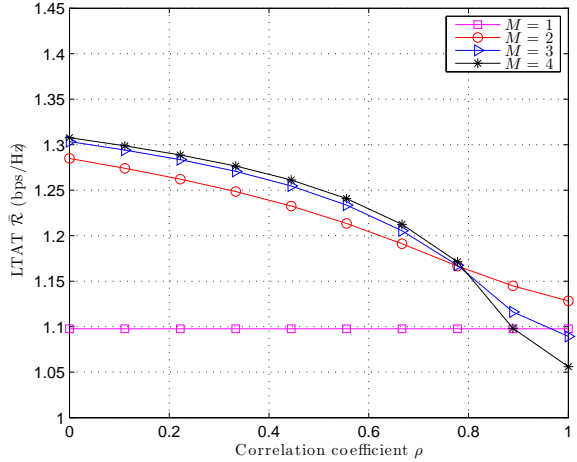}\\
  \caption{LTAT $\bar {\mathcal R}$ versus correlation coefficient. ($\gamma_T=7$dB)}\label{fig_avg_rate_rho}
\end{figure}

The impact of time correlation on the performance of HARQ-IR including the outage probability, the average number of transmissions and the LTAT is now investigated, and the results are shown in Fig. \ref{fig_out_pro_rho}-\ref{fig_avg_rate_rho}, respectively. Notice that for fully correlated fading channels with $\rho=1$, the outage probability is obtained as (\ref{eqn:rho_1_outage}). Correspondingly, the average number of transmissions $\bar {\mathcal N}$ and the LTAT $\bar {\mathcal R}$ can be obtained by putting (\ref{eqn:rho_1_outage}) into (\ref{eqn_mean_trans_num}) and (\ref{eqn_ltat_asym}), respectively.

The outage probability versus the time correlation coefficient under various $\gamma_T$ and $M$ is shown in Fig. \ref{fig_out_pro_rho}. It is readily seen that the outage probability increases with time correlation coefficient $\rho$. Specifically, for the case with $M=4$ and $\gamma_T=7~\rm{dB}$, $P_{out}^{IR}(M)$ increases from $3*10^{-4}$ to $8*10^{-2}$ when $\rho$ increases from $0$ to $1$. It indicates that time correlation does cause negative impact on the outage performance. Additionally, it is observed that the gap between the outage probabilities for two different $M$ becomes narrower when $\rho$ gets higher. In other words, under highly correlated channels, further increase of the number of transmissions will only lead to slight improvement on the outage probability.

In Fig. \ref{fig_avg_num_rho}, the average number of transmissions $\bar {\mathcal N}$ versus correlation coefficient $\rho$ is plotted. It can be seen that $\bar {\mathcal N}$ increases with $\rho$ for $M>2$. This is also because of the negative impact of the correlation in the channels. When the time correlation increases, outage probability will increase, thus more HARQ rounds are required to successfully deliver a message. Notice that when $M=1,2$, the average number of transmissions $\bar {\mathcal N}$ is irrelevant to the time correlation, which directly follows from (\ref{eqn_mean_trans_num}).

With respect to the effect of time correlation on the LTAT, similarly we can find that the LTAT $\bar {\mathcal{R}}$ decreases with the increase of $\rho$ as shown in Fig. \ref{fig_avg_rate_rho}. For instance, the LTAT $\bar {\mathcal{R}}$ for the case with $M=4$ decreases from $1.30~\rm{bps/Hz}$ to $1.05~\rm{bps/Hz}$ as $\rho$ increases from $0$ to $1$. Interestingly, it can also be easily observed that the LTAT shows opposite trends with the increase of the number of transmissions under different correlation regions. More specifically, under low-to-median correlation, the LTAT is improved when the number of transmissions increases. However, when the time correlation is high, additional transmission causes degradation of the LTAT when $M \ge 2$. This is due to the twofold impact of increasing the number of transmissions. On one hand, the increase of the number of transmissions would decrease the outage probability. On the other hand, It would also increase the average number of transmissions. The first impact dominates under low-to-median correlation, while the second one dominates under high correlation. Therefore, from the LTAT's point of view, more transmissions may not be better when time correlation is high.

\subsection{Optimum Rate Design}
As defined in (\ref{eqn_ltat_asym}), the LTAT is a complicated function of the transmission rate $\mathcal{R}$ due to the implicit involvement of $\mathcal{R}$ in outage probability and the average number of transmissions. To maximize the LTAT, the rate should be properly designed. Mathematically, the problem of optimal rate design can be formulated as
\begin{equation}\label{eqn_op2}
\begin{aligned}
& \underset{\mathcal R}{\text{min}}
& & \bar {\mathcal{R}} = \frac{{\mathcal{R}\left( {1 - P_{out}^{IR}\left( M \right)} \right)}}{{\bar {\mathcal{N}} }} \\
& \text{s.t.}
& & P_{out}^{IR}\left( M \right) \le \varepsilon.
\end{aligned}
\end{equation}
With the approximation in (\ref{eqn:appro_acmu}) where $\kappa_i$, ${{\zeta }}$, and ${{\theta }}$ are irrelevant to the rate $\mathcal{R}$, the optimal rate can be easily solved by using optimization tools.

\begin{figure}
  \centering
  \includegraphics[width=3.5in]{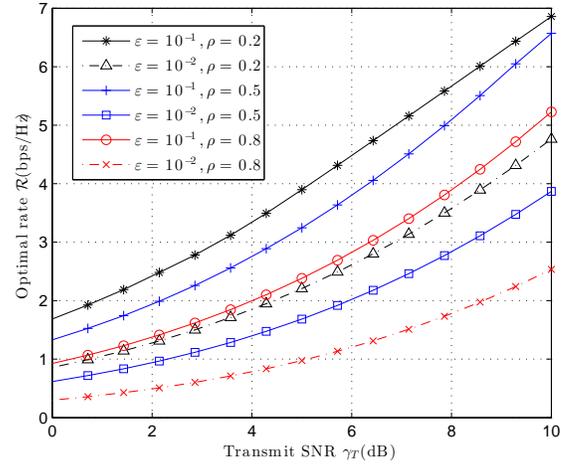}\\
  \caption{Optimum transmission rate $R$ versus transmit SNR $\gamma_T$ for different target outage probability $\varepsilon$ for $M=4$.}\label{fig_optimized_rate}
\end{figure}

Fig. \ref{fig_optimized_rate} shows the optimal rate given various outage constraints $\varepsilon$ under different correlation scenarios. It is clear that the optimal transmission rate increases with the transmit SNR and the allowable outage probability $\varepsilon$. For examples, the optimum rate $\mathcal R$ increases by $3.2~\rm{bps/Hz}$ when the transmit SNR $\gamma_T$ is increased from $0~\rm{dB}$ to $10~\rm{dB}$ for the case with $\varepsilon = 10^{-2}$ and $\rho=0.5$. It also increases from $3.87~\rm{bps/Hz}$ to $6.57~\rm{bps/Hz}$ when the allowable outage probability $\varepsilon$ increases from $0.01$ to $0.1$ for the case with $\rho=0.5$ and $\gamma_T=10~\rm{dB}$. On the other hand, time correlation of the channels has negative effect on the optimal rate.

\begin{figure}
  \centering
  \includegraphics[width=3.5in]{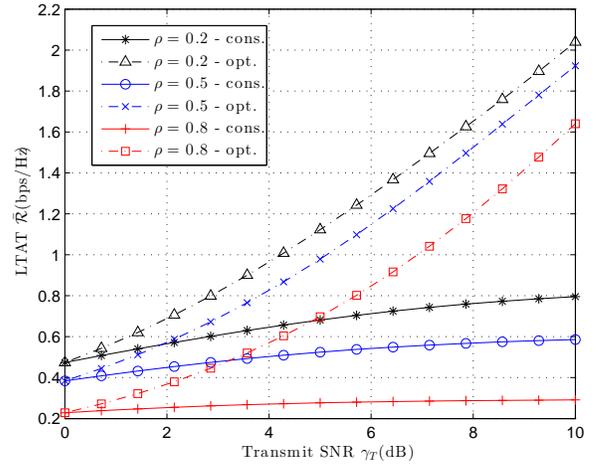}\\
  \caption{LTAT $\bar {\mathcal R}$ for the schemes with optimum rate design and a constant rate.}\label{fig_optimized_avg_trans_rate}
\end{figure}

To further investigate the improvement of LTAT through optimal rate design, the LTAT $\bar {\mathcal R}$ versus transmit SNR $\gamma_T$ for the schemes with optimal rate design and a constant rate (which is set as the optimal rate corresponding to the case with $\gamma_T = 0~\rm{dB}$), is depicted in Fig. \ref{fig_optimized_avg_trans_rate} by setting $\varepsilon = 0.01$ and $M=4$. Apparently, notable improvement of LTAT can be observed through optimal rate design and the contribution of the optimal rate design becomes more significant when the transmit SNR gets higher.
\section{Conclusions}
\label{sec_con}
Performance of HARQ-IR scheme operating over time-correlated Rayleigh fading channels has been analyzed in this paper. By using polynomial fitting technique, the PDF of the accumulated mutual information has been derived, which enables the derivation of outage probability, average number of transmission and LTAT in closed-forms. It has been found that the outage probability can be written as a weighted sum of outage probabilities corresponding to a number of Gamma RVs. Moveover, diversity order has been analyzed and it has been revealed that full diversity can be achieved even under time-correlated fading channels. The impact of time correlation in the channels has also been investigated. It has been demonstrated that time correlation has negative effect on the performance. Under highly correlated channels, more transmissions would not necessarily lead to a higher LTAT and a few transmissions may be sufficient. Finally, the analytical results have enabled the optimal design of HARQ-IR scheme and optimal rate design has been particularly discussed to demonstrate the significance of our analytical results on HARQ-IR.
\appendices
\section{Derivation of moments ${\mathcal M}(i)$}\label{dm}
By definition, the $i$th moment of ${I_K^{IR}}$ is expressed as
\begin{multline}\label{eqn_mom_exp}
{\cal M}\left( i \right) = \int_{{x_1} = 0}^\infty   \cdots \int_{{x_K} = 0}^\infty  {{\left( {\sum\limits_{k = 1}^K {{{\log }_2}\left( {1 + {x_k}} \right)} } \right)}^i} \\
\times {f_{{{\bf{\gamma }}_{1:K}}}}\left( {{x_1}, \cdots ,{x_K}} \right)d{x_1} \cdots d{x_K}  .
\end{multline}
By substituting (\ref{eqn_joint_pdf_X}) into (\ref{eqn_mom_exp}), it follows that
\begin{multline}\label{eqn_m_i_org}
{\cal M}\left( i \right) = \prod\limits_{k = 1}^K {\frac{1}{{2{\sigma _k'}{^2}\left( {1 - {\lambda _k}^2} \right)}}}   \int\limits_{t = 0}^\infty {{\rm{e}}^{ - \left( {1 + \sum\limits_{k = 1}^K {\frac{{{\lambda _k}^2}}{{1 - {\lambda _k}^2}}} } \right)t}} \\
 \times {\int\limits_0^\infty  { \cdots \int\limits_0^\infty  {{{\left( {\sum\limits_{k = 1}^K {{{\log }_2}\left( {1 + {x_k}} \right)} } \right)}^i}\prod\limits_{k = 1}^K {e^{ - \frac{{{x_k}}}{{2{\sigma_k'}{^2}\left( {1 - {\lambda _k}^2} \right)}}}}} } } \\
\times {}_0{F_1}\left( {;1;\frac{{{x_k}{\lambda _k}^2t}}{{2{\sigma _k'}{^2}\left( {1 - {\lambda _k}^2} \right)^2}}} \right)d{x_1} \cdots d{x_k} dt.
\end{multline}
Using binomial expansion, (\ref{eqn_m_i_org}) can be rewritten as
\begin{multline}\label{eqn_fur_mom}
{{\cal M}\left( i \right) = \prod\limits_{k = 1}^K {\frac{1}{{2{\sigma _k'}{^2}\left( {1 - {\lambda _k}^2} \right)}}} \sum\limits_{{i_1} = 0}^i {\sum\limits_{{i_2} = 0}^{i - {i_1}} { \cdots \sum\limits_{{i_{K - 1}} = 0}^{i - \sum\limits_{l = 1}^{K - 2} {{i_l}} }  } C_i^{{i_1}}} }\\
 \times C_{i - {i_1}}^{{i_2}} \cdots C_{i - \sum\limits_{l = 1}^{K - 1} {{i_l}}}^{{i_{K}}} \int\limits_{t = 0}^\infty {{\rm{e}}^{ - \left( {1 + \sum\limits_{k = 1}^K {\frac{{{\lambda _k}^2}}{{1 - {\lambda _k}^2}}} } \right)t}} \\
 \times \prod\limits_{k = 1}^K \int\limits_0^\infty \left({{\log }_2}\left( {1 + {x_k}} \right)\right)^{{i_k}}{e^{ { - \frac{{{x_k}}}{{2{\sigma _k'}{^2}{{\left( {1 - {\lambda _k}^2} \right)}}}}} }}\\
\times  {  {{}_0{F_1}\left( {;1;\frac{{{x_k}{\lambda _k}^2t}}{{2{\sigma _k'}{^2}\left( {1 - {\lambda _k}^2} \right)^2}}} \right)} d{x_k}}  dt.
\end{multline}
Then making change of variables yields
\begin{multline}\label{eqn:fur_mom_fur}
{\cal M}\left( i \right) = \frac{{i!}}{{1 + \sum\limits_{k = 1}^K {\frac{{{\lambda _k}^2}}{{1 - {\lambda _k}^2}}} }}\sum\limits_{\sum\limits_{l=1}^{K}{i_l}=i,{i_l} \ge 0} {\frac{1}{{{i_1}!{i_2}! \cdots {i_K}!}}  } \\
\times \int\limits_{t = 0}^\infty  {{{\rm{e}}^{ - t}}\prod\limits_{k = 1}^K {\int\limits_0^\infty  {{e^{ - y}}{{\log }_2}^{{i_k}}\left( {1 + {w_k}y} \right){}_0{F_1}\left( {;1;{\varpi _k}yt} \right)} dy} } dt
\end{multline}
where ${\varpi _k} = \frac{{\frac{{{\lambda _k}^2}}{{1 - {\lambda _k}^2}}}}{{1 + \sum\limits_{l = 1}^K {\frac{{{\lambda _l}^2}}{{1 - {\lambda _l}^2}}} }}$, ${w_k} = 2{\sigma _k'}^2\left( {1 - {\lambda _k}^2} \right)$.

By applying Gaussian quadrature into (\ref{eqn_fur_mom}) \cite{Germund2008numerical}, it produces
\begin{multline}\label{eqn_mom_app_gauss}
{\mathcal M}\left( i \right) \approx \frac{{i!}}{{1 + \sum\limits_{k = 1}^K {\frac{{{\lambda _k}^2}}{{1 - {\lambda _k}^2}}} }}\sum\limits_{\sum\limits_{l=1}^K{i_l}=i,{i_l} \ge 0} {\frac{1}{{{i_1}!{i_2}! \cdots {i_K}!}}  } \\
\times \int\limits_{t = 0}^\infty  {{{\mathop{\rm e}\nolimits} ^{ - t}}\prod\limits_{k = 1}^K {\sum\limits_{{q_k} = 1}^{{N_Q}} {{\varrho _{{q_k}}}} {{\log }_2}^{{i_k}}\left( {1 + {w_k}{\xi _{{q_k}}}} \right){}_0{F_1}\left( {;1;{\varpi _k}{\xi _{{q_k}}}t} \right)} } dt\\
 = \frac{{i!}}{{1 + \sum\limits_{k = 1}^K {\frac{{{\lambda _k}^2}}{{1 - {\lambda _k}^2}}} }}\sum\limits_{\sum\limits_{l=1}^K{i_l}=i,{i_l} \ge 0} {\frac{1}{{{i_1}!{i_2}! \cdots {i_K}!}} \sum\limits_{{q_k} \in \left[ {1,{N_Q}} \right],k \in \left[ {1,K} \right]} } \\
 {\prod\limits_{k = 1}^K {{\varrho _{{q_k}}}{{\log }_2}^{{i_k}}\left( {1 + {w_k}{\xi _{{q_k}}}} \right)} \int\limits_{t = 0}^\infty  {{{\mathop{\rm e}\nolimits} ^{ - t}}\prod\limits_{k = 1}^K {{}_0{F_1}\left( {;1;{\varpi _k}{\xi _{{q_k}}}t} \right)} } dt}
\end{multline}
where $N_Q$ is the quadrature order, and the weights $\varrho_{q_k}$ and abscissas ${{\xi _{q_k}}}$ are tabulated in \cite[Table 25.9]{abramowitz1972handbook}. The residue error becomes negligible if $N_Q$ is sufficiently large. By using the following formula
\begin{align}\label{eqn_use_int}
&{\int\limits_{z = 0}^\infty  {{e^{ - z}}\prod\limits_{l = 1}^K {_0{F_1}\left( {;1;{a_l}z} \right)} } dz} \notag \\
 &= \int\limits_{z = 0}^\infty  {{e^{ - z}}\prod\limits_{l = 1}^K {\frac{1}{{2\pi {\rm j}}}\int\limits_{{{\cal C}_l}} {\frac{{\Gamma \left( s_l \right)}}{{\Gamma \left( {1 - s_l} \right)}}{{\left( { - {a_l}z} \right)}^{ - {s_l}}}d{s_l}} } } dz \notag\\
 &= {{\left( {\frac{1}{{2\pi {\rm j}}}} \right)}^K}\int\limits_{{{\cal C}_1}} { { \cdots \int\limits_{{{\cal C}_K}} {\frac{{\Gamma \left( {1 - \sum\limits_{l = 1}^K {{s_l}} } \right)\prod\limits_{l = 1}^K {\Gamma \left( {{s_l}} \right)} }}{{\prod\limits_{l = 1}^K {\Gamma \left( {1 - {s_l}} \right)} }}} } } \notag \\
 &\times {{\left( { - {a_1}} \right)}^{ - {s_1}}} \cdots {{\left( { - {a_K}} \right)}^{ - {s_K}}} d{s_1} \cdots d{s_K} \notag \\
 &= \Psi _2^{\left( K \right)}\left( {1;\underbrace {1,1, \cdots ,1}_{K - terms};{a_1},{a_2}, \cdots ,{a_K}} \right)
\end{align}
where ${\rm j}=\sqrt{-1}$, and $\Psi _2^{\left( K \right)}(;;)$ denotes confluent form of Lauricella hypergeometric function \cite[Definition A.20]{mathai2009h}, \cite{saigo1992some}, the final expression for (\ref{eqn_mom_app_gauss}) then follows as (\ref{eqn_M_fin_sec}).

\section{Proof of Invertibility of Matrix $\bf{A}$}
\label{app:proof_inver_A}
To complete the proof, it suffices to show that $\bf{A}$ is a positive definite matrix. For arbitrary $(N+1) \times 1$ real vector $\bf u$, a function $\bf{u^{\rm T}}{\bf A}{{\bf{u }}}$ can be derived as
\begin{equation}\label{eqn:quad_form}
{\bf{u^{\rm T} }}{\bf A}{{\bf{u }}} = \int_{ 0 }^\infty  {\varphi\left( x \right)\left({{\bf{u}}^{\rm{T}}\boldsymbol{\mathcal{P}}(x)}\right)^2dx}.
\end{equation}
It is straightforward that $\int_{ 0 }^\infty  {\varphi\left( x \right)\left({{\bf{u}}^{\rm{T}}\boldsymbol{\mathcal{P}}(x)}\right)^2dx} \ge 0$ and the equality holds if and only if $\bf{u}=\bf{0}$. It follows that $\bf{A} \succ 0$, thus $\bf{A}$ is non-singular, that is, $\bf{A}$ is invertible.

\section{Proof of Property \ref{pro:irr}}
\label{app:proof_irre_to_pn}
The proof is completed by contradiction. First, denote two measures as $d\mu _1 (x)$ and $d\mu _2 (x)$, and their corresponding monic orthogonal polynomials as $\boldsymbol {\mathcal {P}}_1(x)$ and  $\boldsymbol {\mathcal {P}}_2(x)$, respectively. As aforementioned, given the measure $d\mu (x)$, the optimal solution to the problem of (\ref{eqn_op}) is unique. We then represent the unique optimal solutions corresponding to the two measures $d\mu _1 (x)$ and $d\mu _2 (x)$ as $\boldsymbol{\eta}_1$ and $\boldsymbol{\eta}_2$, respectively. Meanwhile, the corresponding optimal polynomials are denoted as $\hat \psi _N^{(1)}(x)={\boldsymbol{\eta}_1}^{\rm T}\boldsymbol {\mathcal {P}}_1(x)$ and $\hat \psi _N^{(2)}(x)={\boldsymbol{\eta}_2}^{\rm T}\boldsymbol {\mathcal {P}}_2(x)$, while the corresponding minimal MSEs are ${{\mathcal S}_{mse}}({\boldsymbol{\eta }}_1|d\mu_1 (x),N)$ and ${{\mathcal S}_{mse}}({\boldsymbol{\eta }}_2|d\mu_2 (x),N)$.


Assume that ${{\mathcal S}_{mse}}({\boldsymbol{\eta }}_1|d\mu_1 (x),N) \ne {{\mathcal S}_{mse}}({\boldsymbol{\eta }}_2|d\mu_2 (x),N)$, which implies that $\hat \psi_N^{(1)}(x) \ne \hat \psi_N^{(2)}(x)$. Without loss of generality, we consider the case with ${{\mathcal S}_{mse}}({\boldsymbol{\eta }}_1|d\mu_1 (x),N) > {{\mathcal S}_{mse}}({\boldsymbol{\eta }}_2|d\mu_2 (x),N)$. It is known that any polynomial $\hat \psi_N(x)\in \mathbb {P}_N$ has a unique coordinate $\boldsymbol{\eta}$ regarding to a given basis of orthogonal polynomials. Thus by taking $\boldsymbol {\mathcal {P}}_1(x)$ as a basis, we have $\hat \psi_N^{(2)}(x)={\boldsymbol{\eta}_3}^{\rm T}\boldsymbol {\mathcal {P}}_1(x)$, where ${\boldsymbol{\eta }}_3 $ is the unique coordinate associated with $\hat \psi_N^{(2)}(x)$ on the basis of $\boldsymbol {\mathcal {P}}_1(x)$. Considering the unique optimality of ${\boldsymbol{\eta }}_1 $ given the measure $d\mu _1 (x)$, we have ${{\mathcal S}_{mse}}({\boldsymbol{\eta }}_1|d\mu_1 (x),N) < {{\mathcal S}_{mse}}({\boldsymbol{\eta }}_3|d\mu_1 (x),N)={{\mathcal S}_{mse}}({\boldsymbol{\eta }}_2|d\mu_2 (x),N)$. This is contradictory with our assumption. Therefore,  ${{\mathcal S}_{mse}}({\boldsymbol{\eta }}_1|d\mu_1 (x),N) = {{\mathcal S}_{mse}}({\boldsymbol{\eta }}_2|d\mu_2 (x),N)$ and $\hat \psi_N^{(1)}(x) = \hat \psi_N^{(2)}(x)$. Then the proof follows.


\bibliographystyle{IEEEtran}
\bibliography{gamma}
\begin{IEEEbiography}[{\includegraphics[width=1in,height=1.25in,clip,keepaspectratio]{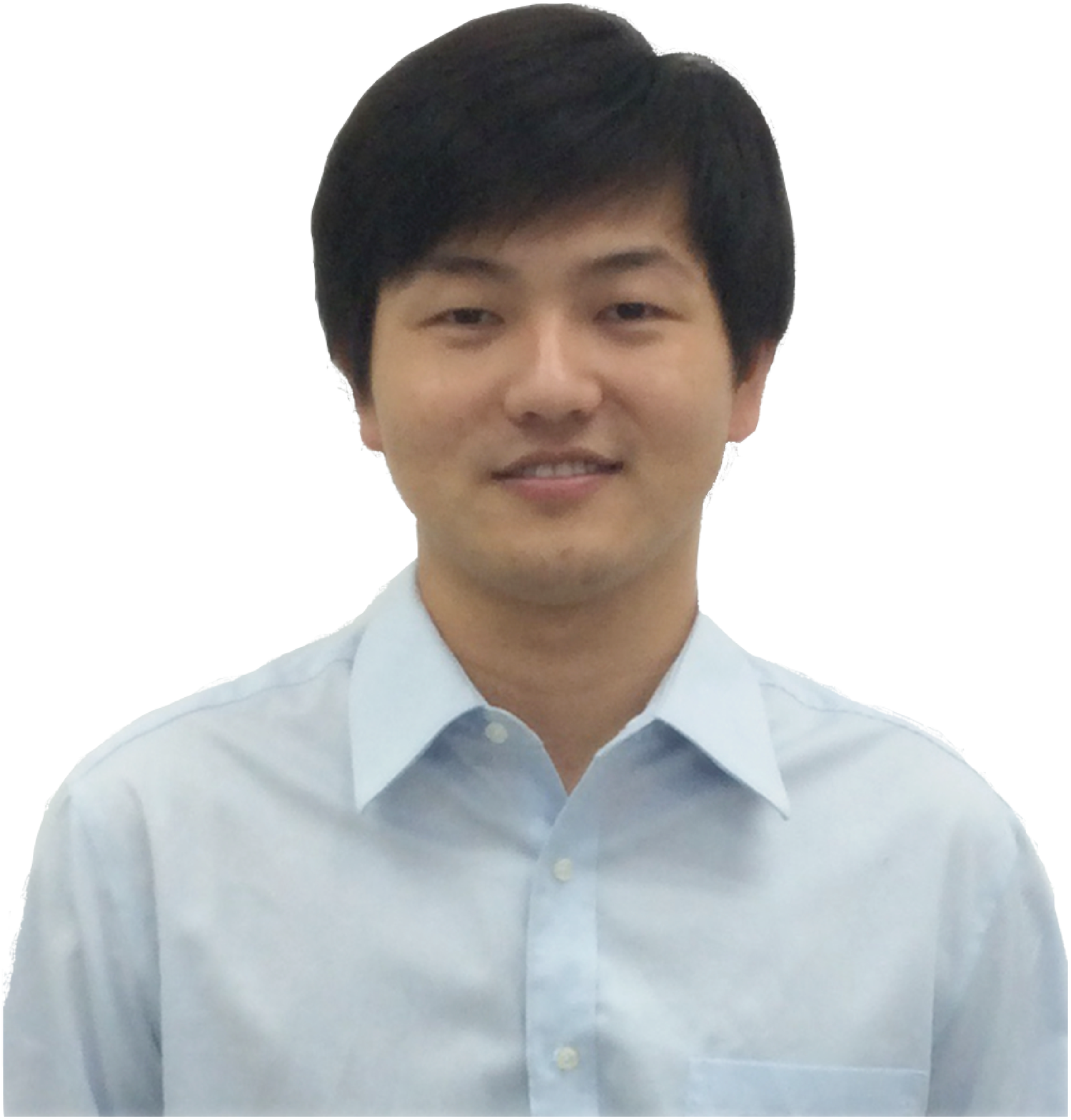}}]{Zheng Shi}
received the B.S. degree in communication engineering from Anhui Normal University, China, in 2010 and the M.S. degree in communication and information system from Nanjing University of Posts and Telecommunications (NUPT), China, in 2013. Since Sep. 2013, he has been a Ph.D. student in Department of Electrical and Computer Engineering, University of Macau, Macao. His research interests include hybrid automatic repeat request (HARQ) protocols, cooperative communications, full-duplex communications, massive MIMO, and heterogeneous wireless networks.
\end{IEEEbiography}

\begin{IEEEbiography}[{\includegraphics[width=1in,height=1.25in,clip,keepaspectratio]{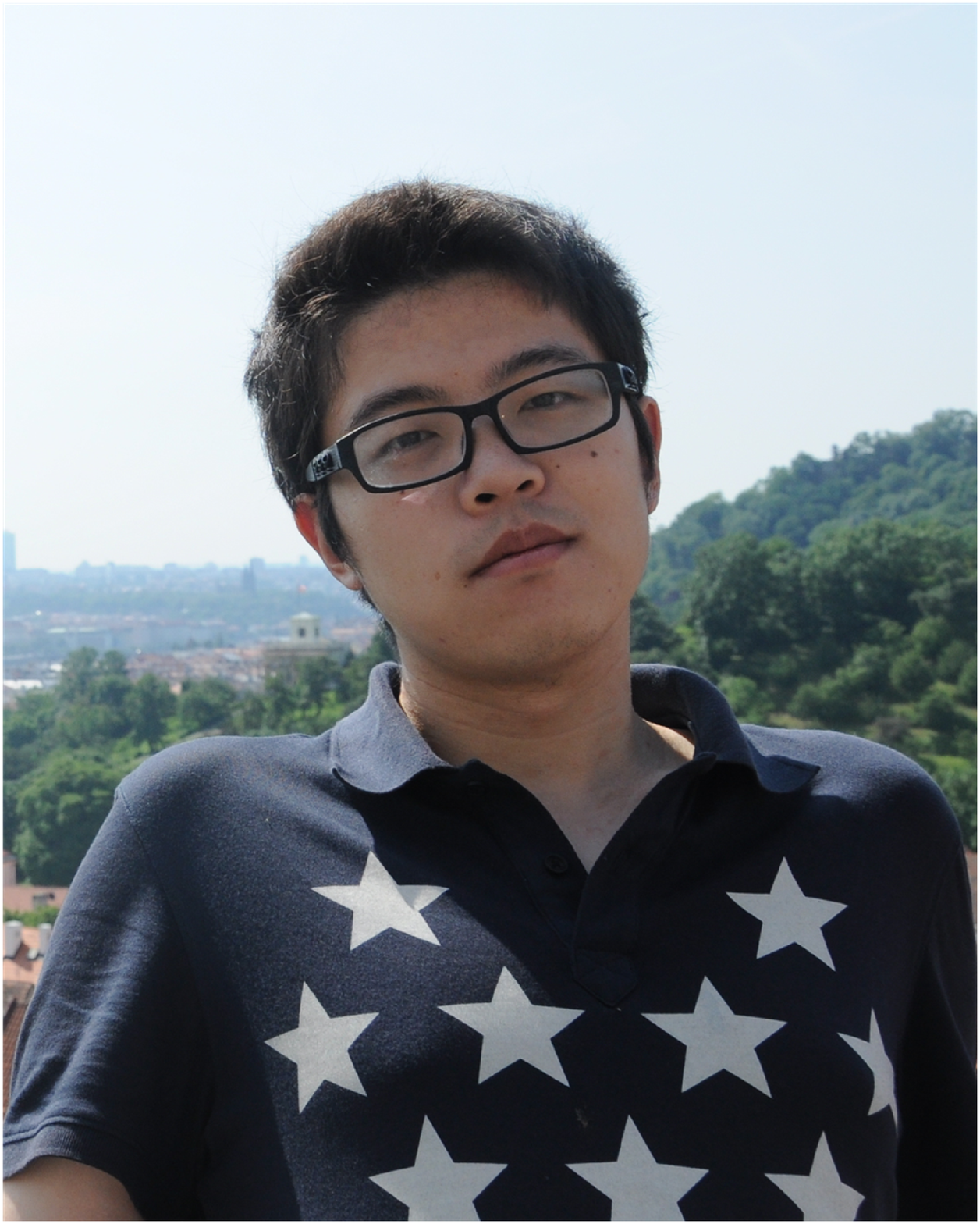}}]{Haichuan Ding} is currently a Ph.D. student at the University of Florida. He received the B.Eng. and M.S. degrees in electrical engineering from Beijing Institute of Technology (BIT), Beijing, China, in 2011 and 2014, respectively. From 2012 to 2014, he was with the Department of Electrical and Computer Engineering, University of Macau, as a visiting student. During his M.S. studies, he mainly worked on the analysis of HARQ techniques using the tools of stochastic geometry. His current research is focused on cognitive radio networks and security and privacy in distributed systems.
\end{IEEEbiography}

\begin{IEEEbiography}[{\includegraphics[width=1in,height=1.25in,clip,keepaspectratio]{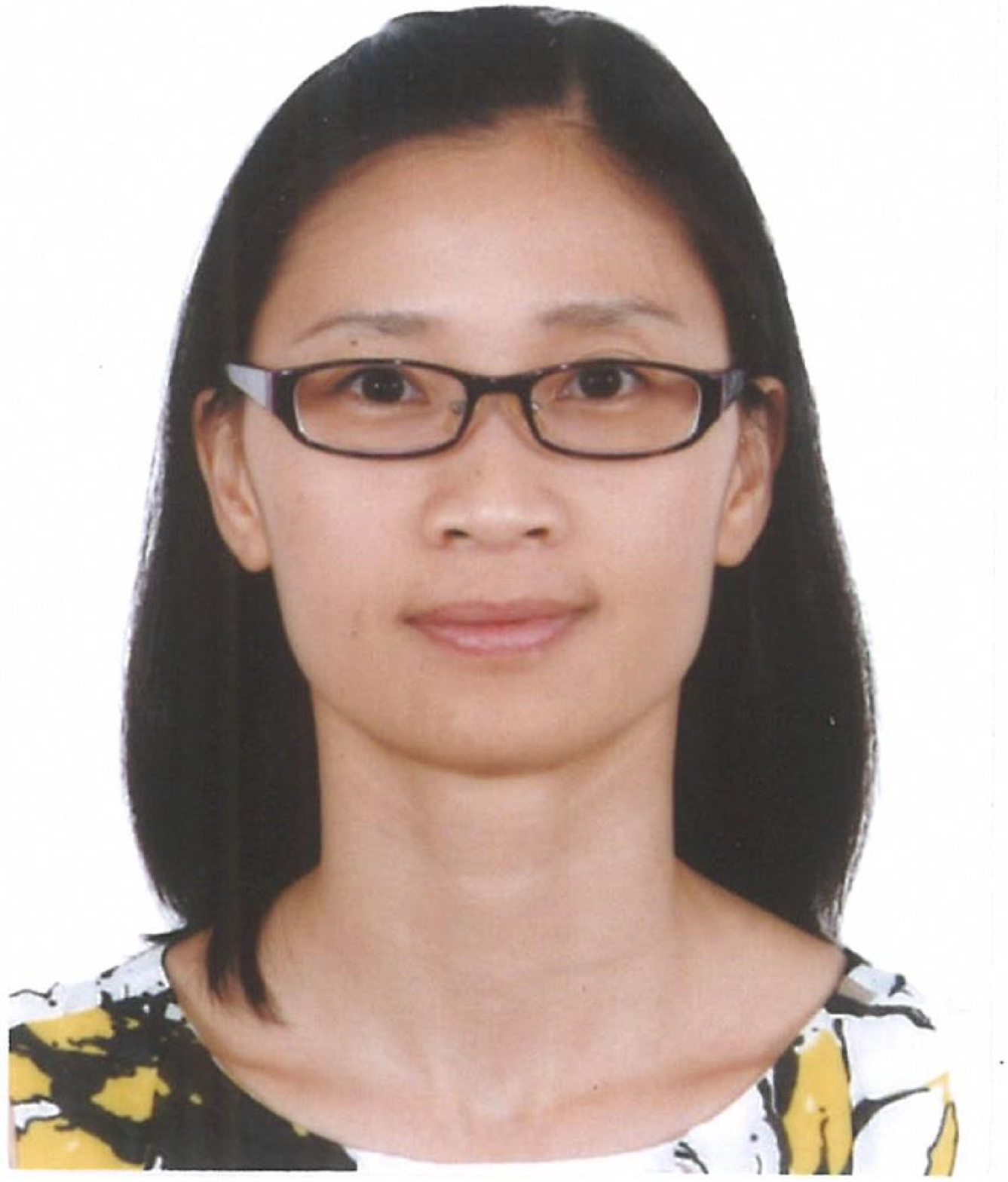}}]{Shaodan Ma}
received her double Bachelor degrees in Science and Economics, and her Master degree of Engineering, from Nankai University, Tianjin, China. She obtained her Ph. D. degree in electrical and electronic engineering from the University of Hong Kong, Hong Kong, in 2006. After graduation, she joined the University of Hong Kong as a Postdoctoral Fellow. Since August 2011, she has been with the University of Macau and is now an Associate Professor there. She was a visiting scholar in Princeton University in 2010 and is currently an Honorary Assistant Professor in the University of Hong Kong.

Her research interests are in the general areas of signal processing and communications, particularly, transceiver design, resource allocation and performance analysis.
\end{IEEEbiography}
\begin{IEEEbiography}[{\includegraphics[width=1in,height=1.25in,clip,keepaspectratio]{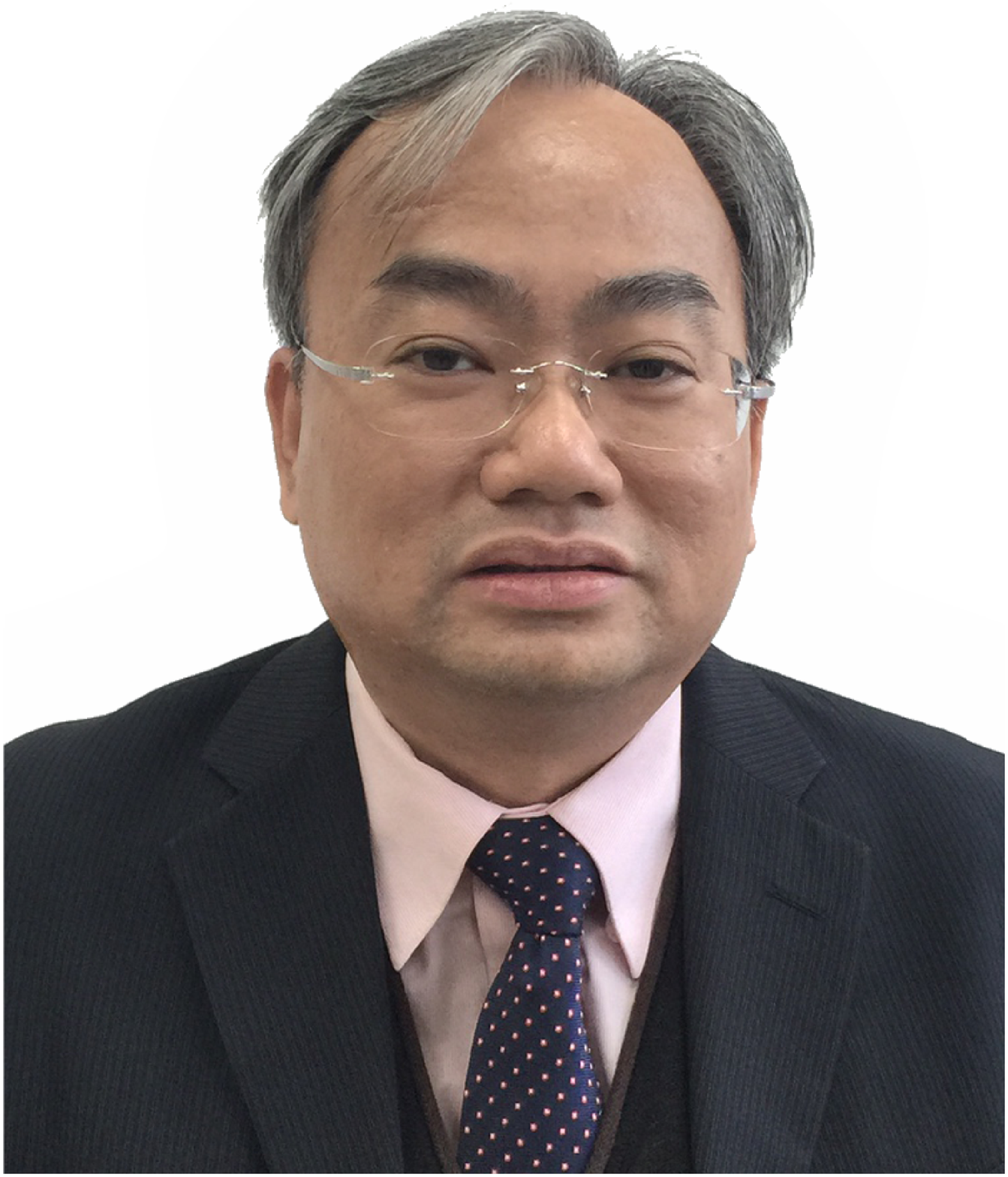}}]{Kam-Weng Tam}
(S'91-M'01-SM'05) received the B.Sc. and joint Ph.D. degrees in electrical and electronics engineering from the University of Macau, Taipa, Macao, China, and the University of Macau and Instituto Superior Técnico (IST), Technical University of Lisbon, Lisbon, Portugal, in 1993 and 2000, respectively.

From 1993 to 1996, he was with the Instituto de Engenharia de Sistemas e Computadores (INESC), Lisbon, Portugal, where he participated in research and development on a broad range of applied microwave technologies for satellite communication systems. From July 2000 to December 2001, he was the Director of the Instituto de Engenharia de Sistemas e Computadores (INESC)-Macau. In 2001, he cofounded the microelectronic design house Chipidea Microelectrónica, Macau, China, where until 2003 he was the General Manager. Since 1996, he has been with the University of Macau, where he is currently a Professor and the Associate Dean (Research and Graduate Studies) with the Faculty of Science and Technology. He has authored or coauthored over 100 journal and conference papers. His research interests have concerned multifunctional microwave circuits, RFID, UWB for material analysis and terahertz technology.

Dr. Tam was interim secretary for the establishment of the Macau Section in 2003. He supervised two IEEE Microwave Theory and Techniques Society (IEEEMTT-S) Undergraduate Scholarship recipients in 2002 and 2003. He was founder of the IEEE Macau AP/MTT Joint Chapter in 2010 and was chair in 2011–2012. He was a member of the organizing committees of 21 international and local conferences including co-chair of APMC2008, co-chair of the Technical Program, IEEE MTT-S International Microwave Workshop Series on Art of Miniaturizing RF and Microwave Passive Components (2008), and co-chair of ISAP2010.
\end{IEEEbiography}
\end{document}